\documentclass[aps, prd, showpacs ]{revtex4}

\usepackage{amsmath,amsfonts,amscd,amssymb,epsf}
\usepackage[small]{caption}

\newcommand{\pa}{\partial}

\newcommand{\vep}{\varepsilon}

\begin{document}

 \title{ Casimir effect between two spheres at small separations}

 \author{L. P. Teo}\email{LeePeng.Teo@nottingham.edu.my}
\address{Department of Applied Mathematics, Faculty of Engineering, University of Nottingham Malaysia Campus, Jalan Broga, 43500, Semenyih, Selangor Darul Ehsan, Malaysia.}
\begin{abstract}
We consider the Casimir interaction between two spheres at  zero and finite temperature, for both scalar fields and electromagnetic fields. Of particular interest is the asymptotic expansions of the Casimir free energy when the distance between the spheres is small. The scenario where one sphere is inside the other is discussed in detail. At zero temperature, we compute analytically the leading and the next-to-leading order terms  from the functional determinant representation of the Casimir energy. As expected, the leading order term agrees with the proximity force approximation. The results for the next-to-leading order terms are new. In the limit where the radius of the outer sphere goes to infinity,   the results for the sphere-plane geometry are reproduced. At finite temperature,   the leading order term is computed and it is found to agree completely with the proximity force approximation in the medium and high temperature regions. For the scenario where two spheres are outside each other, analogous results are obtained. In the case of Dirichlet boundary conditions on both spheres, the  next-to-leading order term of the zero temperature Casimir energy is found to agree with that computed recently using derivative expansion.

\end{abstract}
\maketitle

\section{Introduction}

In recent years, the Casimir interactions between two compact objects have been under active investigation. Different methods have been developed to compute the Casimir interactions beyond the accuracies afforded by the proximity force approximations. Using functional determinant or multiple scattering approach \cite{1,2,3,4,5,6,7,8,9}, the Casimir interaction between two objects can be written in terms of the scattering matrices of each object and the translation matrices which relate the different bases for computing the scattering of each object.
In principal, the functional representation of the Casimir interaction allows one to numerically compute the exact magnitude of the Casimir interaction at any separation. However, the matrices involved are often infinite matrices and truncations are required, which at smaller separation, pose a big challenge since the size of the matrices required for accurate computation is inversely proportional to the distance between  the objects. On the analytical side, the situation is the same. It is easy to obtain the large separation asymptotic expansion of the Casimir interaction   from a finite number of terms in the matrices. However, to obtain the small separation asymptotic expansions, all the terms have to be taken into account.

In the pioneering work \cite{1}, a perturbation method have been developed to compute the small separation leading order and   next-to-leading order terms of the zero temperature Casimir interaction between a cylinder and a plane. This method was later extended to the sphere-plane \cite{16,17,22} and  cylinder-cylinder \cite{18} configurations, as well as the finite temperature case \cite{15,26}.
Very recently, a new method called derivative expansion has been developed for computing the small separation next-to-leading order term of the Casimir interaction \cite{24,20,25}. While the method of \cite{1,16,17,22,18,15,26} derives the asymptotic expansion from the exact functional determinant representation of the Casimir interaction, the methods of \cite{24,20,25} are completely different. In \cite{24}, the authors considered the scalar interaction between an object imposed with Dirichlet boundary conditions and a Dirichlet plane. They computed the asymptotic expansion by performing a derivative expansion on the field in the path integral representation of the Casimir free energy. The leading order term coincides with the proximity force approximation and a general expression for the next-to-leading order term was derived. On the other hand, in \cite{20,25}, the next-to-leading order term is computed based on a postulate that the Casimir free energy   have a local expansion of a certain form involving the gradient of the height profiles of the objects.

In this article, we consider the   Casimir interaction between two spheres, for both scalar fields and electromagnetic fields. This is one of the most popular configuration under consideration \cite{2,4,5,6,7,9,21,12,27}.  The functional determinant representations of the Casimir interactions have been obtained in various works \cite{2,4,5,6,7,9}, which were used to compute the large separation asymptotic expansions   analytically. For  small separations which are more experimentally relevant,  it has been claimed that the first two leading terms can be computed using the derivative expansion method developed in \cite{20}. Since  this method   is based on a postulate that has not yet been proven, we find that it is necessary to compute the asymptotic expansions from the exact representation of the Casimir interaction. This task is undertaken in this paper. We will consider both the cases where one sphere is inside the other, and the two spheres are outside each other.

Throughout this paper, we use units with $\hbar=c=k_B=1$.

\section{Functional Determinant representation for the Casimir free energy}\label{s2}
In this section, we collect the basic formulas for the Casimir free energy between two spheres $A$ and $B$ with radii $r_A$ and $r_B$ respectively, where $r_A\leq r_B$. Let $L$ be the distance between the centers of the spheres and let $d$ be the distance between the spheres.  There are two scenarios:

\begin{enumerate}
\item[$\bullet$] The sphere $A$ is inside the sphere $B$. In this case, $d=r_B-r_A-L$.
\item[$\bullet$] The two spheres are outside each other. In this case, $d=L-r_A-r_B$.
\end{enumerate}

  Starting from  the representation for the zero temperature Casimir energy, the representation for the  Casimir free energy can  be obtained easily using the Matsubara formalism.
The formula for the zero temperature Casimir energy can be derived using the multiple scattering or functional determinant approach \cite{1,2,3,4,5,6,7,8,9} or the mode summation approach \cite{10,11}. For a scalar field $\varphi$, it is given  by
\begin{equation}\label{eq11_29_1}
E_{\text{Cas}}^{\text{XY},T=0}=\frac{1}{2\pi}\int_0^{\infty}d\xi\,\text{Tr}\ln \left(1-M^{\text{XY}}(\xi)\right).
\end{equation}Here X and Y are respectively the boundary conditions on spheres $A$ and $B$. They are equal to  D, N or R for Dirichlet, Neumann or Robin boundary conditions. In the following, when we say Robin boundary conditions, they include Neumann boundary condition as a special case. The trace Tr is the orbital momentum sum:
$$\text{Tr}= \sum_{m=-\infty}^{\infty}\sum_{l=|m|}^{\infty}, $$and $M^{\text{XY}}$ is a matrix  which can be written as a product of four matrices: $$M^{\text{XY}}=T^{A,\text{X}}U^{AB}T^{B,\text{Y}}U^{BA}.$$ The matrices $U^{AB}$ and $U^{BA}$ are called translation matrices. Their elements are given by
\begin{equation*}
U_{l,\tilde{l}}^{\left\{\substack{AB\\BA}\right\}}(\xi)=(-1)^{l+m}\sqrt{\frac{\pi}{2\xi L}}\sum_{l^{\prime\prime}=|l-\tilde{l}|}^{l+\tilde{l}}(\pm 1)^{l^{\prime\prime}}\sqrt{(2l+1)(2\tilde{l}+1)}(2l^{\prime\prime}+1)\begin{pmatrix} l & \tilde{l} & l^{\prime\prime}\\ 0 & 0 & 0\end{pmatrix}
\begin{pmatrix} l & \tilde{l} & l^{\prime\prime}\\ m & -m & 0\end{pmatrix}
Z_{l^{\prime\prime}+1/2}(\xi L),
\end{equation*}which involve $3j$-symbols. $Z_{l^{\prime\prime}+1/2}(\xi L)$ is a modified Bessel function which depends on the relative position of the spheres. If the sphere $A$ is inside sphere $B$, $$Z_{l^{\prime\prime}+1/2}(\xi L)
=I_{l^{\prime\prime}+1/2}(\xi L),$$whereas if the two spheres are outside each other,
 $$Z_{l^{\prime\prime}+1/2}(\xi L)
=K_{l^{\prime\prime}+1/2}(\xi L).$$
The matrices $T^{A,\text{X}}$ and $T^{B,\text{Y}}$, which are related to the scattering matrices, are called transition matrices. They only  depend on the boundary conditions on the sphere $A$ and the sphere $B$ respectively, and they are diagonal matrices.
For $T^{A,\text{X}}$, if the scalar field satisfies   Dirichlet boundary condition $\varphi=0$ on the sphere $A$, then the $(l,l)$-diagonal element  is given by
\begin{equation*}
T^{A,\text{D}}_{l}(\xi)=\frac{ I_{l+1/2}(\xi r_A) }{ K_{l+1/2}(\xi r_A) };
\end{equation*}whereas if the scalar field satisfies the Robin boundary condition $\pa_n\varphi +\alpha_A\varphi=0$ with parameter $\alpha_A$ on the sphere $A$, then
\begin{equation*}
T^{A,\text{R}}_{l}(\xi)=\frac{u_AI_{l+1/2}(\xi r_A)+\xi r_A I_{l+1/2}'(\xi r_A  )}{u_AK_{l+1/2}(\xi r_A)+\xi r_A   K_{l+1/2}'(\xi r_A  )},
\end{equation*}with $u_A=\alpha_A-1/2$. For sphere $B$, if the sphere $A$ is inside sphere $B$, then
\begin{equation*}
T^{B,\text{D}}_{l}(\xi)=\frac{ K_{l+1/2}(\xi r_B) }{ I_{l+1/2}(\xi r_B) }
\quad\text{and}\quad
T^{B,\text{R}}_{l}(\xi)=\frac{u_BK_{l+1/2}(\xi r_B)+\xi r_B   K_{l+1/2}'(\xi r_B  )}{u_BI_{l+1/2}(\xi r_B)+\xi r_B I_{l+1/2}'(\xi r_B  )}
\end{equation*}respectively for Dirichlet boundary condition and Robin boundary condition with parameter $\alpha_B=u_B+1/2$. If the two spheres are outside each other, then $T_l^{B}$ can be obtained from the corresponding $T_l^{A}$ by changing $A$ to $B$.

For electromagnetic field, the functional determinant representation of the zero temperature Casimir energy between two spheres has been derived explicitly   in \cite{9,12} when one sphere is inside the other, and in \cite{9} when the two spheres are outside each other. It is given by
 \begin{equation}\label{eq12_01_1}
E_{\text{Cas}}^{\text{WZ},T=0}=\frac{1}{2\pi}\int_0^{\infty}d\xi\,\text{Tr}\ln \left(1-\mathbb{M}^{\text{WZ}}(\xi)\right).
\end{equation}Here W and Z are respectively the boundary conditions on spheres $A$ and $B$. They are equal to C or P for perfectly conducting or infinitely permeable boundary conditions respectively. The trace Tr is
\begin{equation*}
\text{Tr}=\sum_{m=-\infty}^{\infty}\sum_{l=\max\{1,|m|\}}^{\infty}\text{tr},
\end{equation*}where the trace tr on the right hand side is a trace over $2\times 2$ matrices:
\begin{equation*}
\mathbb{M}_{l,l'}^{\text{WZ}}=\mathbb{T}^{A,\text{W}}_l\sum_{\tilde{l}=0}^{\infty}\mathbb{U}_{l\tilde{l}}^{AB}\mathbb{T}_{\tilde{l}}^{B,\text{Z}}
\mathbb{U}^{BA}_{\tilde{l}l'}.
\end{equation*}Here
\begin{equation*}
\mathbb{U}_{l,\tilde{l}}^{\left\{\substack{AB\\BA}\right\}}=U_{l,\tilde{l}}^{\left\{\substack{AB\\BA}\right\}}\begin{pmatrix}\Lambda_{l\tilde{l}}^{l^{\prime\prime}} & \tilde{\Lambda}_{l\tilde{l}}\\ \tilde{\Lambda}_{l\tilde{l}} &\Lambda_{l\tilde{l}}^{l^{\prime\prime}}\end{pmatrix},
\end{equation*}with
\begin{equation*}
\Lambda_{l\tilde{l}}^{l^{\prime\prime}}=\frac{1}{2}\frac{l^{\prime\prime}(l^{\prime\prime}+1)-l(l+1)-\tilde{l}(\tilde{l}+1)}{\sqrt{l(l+1)\tilde{l}(\tilde{l}+1)}},\hspace{1cm}
 \tilde{\Lambda}_{l\tilde{l}}=\frac{ m\xi L}{\sqrt{l(l+1)\tilde{l}(\tilde{l}+1)}}.
\end{equation*}
The matrices $\mathbb{T}_l^A$ and $\mathbb{T}_l^B$ are diagonal matrices given by
\begin{equation*}
\mathbb{T}_l^{*} =\begin{pmatrix} T^{*,\text{TE}}_l & 0\\ 0& T_l^{*,\text{TM}}  \end{pmatrix}.
\end{equation*}If the sphere *  is imposed with perfectly conducting boundary conditions,
\begin{equation*}
T^{*,\text{C},\text{TE}}_l =T_l^{*,\text{D}},\hspace{1cm}T^{*,\text{C},\text{TM}}_l =-T_l^{*,\text{R}}\Bigr|_{u_*=1/2}.
\end{equation*}If the sphere *  is imposed with infinitely permeable boundary conditions,
\begin{equation*}
T^{*,\text{P},\text{TE}}_l =T_l^{*,\text{R}}\Bigr|_{u_*=1/2},\hspace{1cm}T^{*,\text{P},\text{TM}}_l =-T_l^{*,\text{D}}.
\end{equation*}
It is easy to see that
\begin{equation*}\begin{split}
E_{\text{Cas}}^{\text{CC},T=0}=&E_{\text{Cas}}^{\text{PP},T=0},\\
E_{\text{Cas}}^{\text{CP},T=0}=&E_{\text{Cas}}^{\text{PC},T=0}.
\end{split}\end{equation*}In fact, these are immediate consequences of the electromagnetic duality.

Using Matsubara formalism, one can obtain the finite temperature Casimir free energy from the corresponding zero temperature Casimir energy by changing the integration over the imaginary frequency $\xi$ to summation over the Matsubara frequencies $\xi_p=2\pi pT$.
  For a scalar field, this gives
\begin{equation*}
E_{\text{Cas}}^{\text{XY}} =T\sum_{p=0}^{\infty}\!'\text{Tr}\ln \left(1-M^{\text{XY}}(\xi_p)\right).
\end{equation*}For an electromagnetic field, the matrix $M^{\text{XY}}$ is replaced by the matrix $\mathbb{M}^{\text{WZ}}$. In this formula, the prime $^{\prime}$ on the summation over $p$ means that the term with $p=0$ is weighted with a factor of $1/2$. By taking the limit $T\rightarrow 0$, one recovers the formula for the zero temperature Casimir energy.

\section{Proximity force approximation}\label{s3}

In this section, we use proximity force approximation to find the leading asymptotic behaviors of the Casimir free energy and the Casimir force between two spheres  when the separation between them is small. As in \cite{15}, we will consider three regions:
\begin{enumerate}
\item[(1)] Low temperature region: $dT\ll r_AT\leq r_BT\ll 1$.
\item[(2)] Medium temperature region: $dT\ll 1\ll r_AT \leq r_BT$.
\item[(3)] High temperature region: $1\ll dT\ll r_A T\leq r_BT$.
\end{enumerate}

First we consider the free energy density between a pair of parallel plates. Assume that the plates, plate $A$ and plate $B$, are located at $z=0$ and $z=d$ respectively, and the scalar field is imposed with the Robin boundary conditions:
\begin{equation*}
\left.\beta_A\pa_n\varphi+\alpha_A\varphi\right|_{z=0}=0,\hspace{1cm}\left.\beta_B\pa_n\varphi+\alpha_B\varphi\right|_{z=d}=0.
\end{equation*}Here $\beta_C=0$ if the plate $C$ is imposed with Dirichlet boundary conditions, in which case we take $\alpha_C=1$. Otherwise $\beta_C=1$.
For this configuration, the zero temperature Casimir energy per unit area has been obtained in \cite{13} and confirmed in \cite{5}.
The finite temperature Casimir free energy per unit area was computed in \cite{14} and it is given by
\begin{equation}\label{eq11_24_2}
\mathcal{E}_{\text{Cas}}^{\parallel,\text{XY}}(d)= \frac{T}{2\pi}  \sum_{p=0}^{\infty}\!'\int_{\xi_p}^{\infty}dx\,x\ln\left(1-\frac{(\beta_Ax-\alpha_A)(\beta_Bx-\alpha_B)}
{(\beta_Ax+\alpha_A)(\beta_Bx+\alpha_B)}e^{-2dx}\right).
\end{equation}

To obtain the leading asymptotic behavior of the Casimir free energy between two spheres from the proximity force approximation, one has to integrate $\mathcal{E}_{\text{Cas}}^{\parallel,\text{XY}}(d)$ over one of the spheres, which we take to be sphere $B$, with the distance $d$ being the distance from a point on sphere $B$ to sphere  $A$.

\begin{figure}[h]\centering
\epsfxsize=0.5\linewidth \epsffile{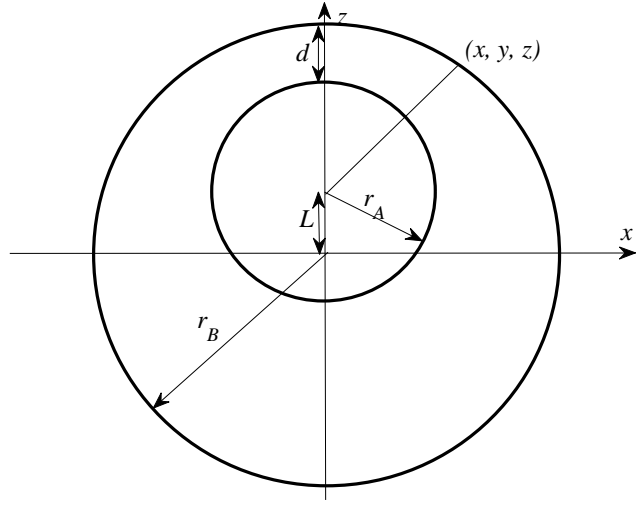} \caption{\label{f1}The cross section of   one sphere inside the  other.}   \end{figure}

First we consider the case where sphere $A$ is inside sphere $B$ (see Fig. \ref{f1}). The sphere $B$ can be parametrized by $(\theta,\phi)$: $x=r_B\sin\theta\cos\phi, y=r_B\sin\theta\sin\phi, z=r_B\cos\theta$, with $0\leq\theta\leq\pi$ and $0\leq\phi<2\pi$. Then the proximity force approximation for the Casimir free energy is
\begin{equation*}
E_{\text{Cas}}^{\text{XY},\text{PFA}}=r_B^2\int_{0}^{2\pi}d\phi\int_0^{\pi}d\theta \sin\theta\mathcal{E}_{\text{Cas}}^{\parallel,\text{XY}}(h(\theta))=2\pi r_B^2\int_0^{\pi}d\theta \sin\theta\mathcal{E}_{\text{Cas}}^{\parallel,\text{XY}}(h(\theta)),
\end{equation*}where $h(\theta)$ is the distance  from the point $(x,y,z)$ on sphere $B$ to sphere $A$ given by
\begin{equation}\label{eq11_24_3}
h(\theta)=\sqrt{r_B^2\sin^2\theta+(r_B\cos\theta-L)^2}-r_A=\sqrt{r_B^2+L^2-2Lr_B\cos\theta}-r_A.
\end{equation}From now on we shall assume that the spheres are eccentric, i.e., $L>0$, and we are interested in the asymptotic behavior of the Casimir free energy when $d\ll L$.
Making a change of variable
$u=h(\theta)$, we find that
\begin{equation}\label{eq11_25_1}
E_{\text{Cas}}^{\text{XY},\text{PFA}}=\frac{2\pi  r_B}{L}\int_{d}^{L+r_B-r_A}du (u+r_A) \mathcal{E}_{\text{Cas}}^{\parallel,\text{XY}}(u).
\end{equation}
Substituting \eqref{eq11_24_2} into \eqref{eq11_25_1}, we have
\begin{equation*}
E_{\text{Cas}}^{\text{XY},\text{PFA}}=\frac{ r_B T} {L}  \sum_{p=0}^{\infty}\!'\int_{d}^{L+r_B-r_A}du (u+r_A) \int_{\xi_p}^{\infty}dx\,x\ln\left(1-\frac{(\beta_Ax-\alpha_A)(\beta_Bx-\alpha_B)}
{(\beta_Ax+\alpha_A)(\beta_Bx+\alpha_B)}e^{-2ux}\right).
\end{equation*}Now making a change of variables $u\mapsto du$ and $x\mapsto x/d$, we find that
\begin{equation*}
E_{\text{Cas}}^{\text{XY},\text{PFA}}=\frac{   r_BT}{d(r_B-r_A-d) }  \sum_{p=0}^{\infty}\!'\int_{1}^{\frac{2r_B-2r_A-d}{d}}du (du+r_A) \int_{ \xi_p d}^{\infty}dx\,x\ln\left(1-\frac{(\beta_Ax-\alpha_A d)(\beta_Bx-\alpha_B d)}
{(\beta_Ax+\alpha_Ad)(\beta_Bx+\alpha_Bd)}e^{-2ux}\right).
\end{equation*}When $d\ll \min\{r_A,r_B, L\}$, this can be approximated by
\begin{equation*}
E_{\text{Cas}}^{\text{XY},\text{PFA}}\sim \frac{  r_A r_BT}{d(r_B-r_A) }  \sum_{p=0}^{\infty}\!'\int_{1}^{\infty}du  \int_{ \xi_p d}^{\infty}dx\,x\ln\left(1-\frac{(\beta_Ax-\alpha_A d)(\beta_Bx-\alpha_B d)}
{(\beta_Ax+\alpha_Ad)(\beta_Bx+\alpha_Bd)}e^{-2ux}\right).
\end{equation*}To proceed further, we need to consider the boundary conditions on the spheres. For XY $=$ DD, $\beta_A=\beta_B=0$ and we find that
$$\frac{(\beta_Ax-\alpha_A d)(\beta_Bx-\alpha_B d)}
{(\beta_Ax+\alpha_Ad)(\beta_Bx+\alpha_Bd)}=1.$$For XY $=$ RD, $\beta_A\neq 0,\beta_B= 0$, and we find that
$$\frac{(\beta_Ax-\alpha_A d)(\beta_Bx-\alpha_B d)}
{(\beta_Ax+\alpha_Ad)(\beta_Bx+\alpha_Bd)}=-\frac{ (\beta_Ax-\alpha_A d)}
{ (\beta_Ax+\alpha_Ad)}\xrightarrow{d\rightarrow 0} -\frac{\beta_A x}{\beta_A x}=-1.$$
Similarly, for XY $=$ DR,
$$\frac{(\beta_Ax-\alpha_A d)(\beta_Bx-\alpha_B d)}
{(\beta_Ax+\alpha_Ad)(\beta_Bx+\alpha_Bd)} \xrightarrow{d\rightarrow 0} -1,$$and for XY $=$ RR,
$$\frac{(\beta_Ax-\alpha_A d)(\beta_Bx-\alpha_B d)}
{(\beta_Ax+\alpha_Ad)(\beta_Bx+\alpha_Bd)} \xrightarrow{d\rightarrow 0} 1.$$
From these, we see that in the cases of DD and RR boundary conditions, the
proximity force approximation of the Casimir free energy gives the same leading behavior:
\begin{equation*}
E_{\text{Cas}}^{\text{PFA}}\sim \frac{  r_A r_BT}{d(r_B-r_A) }  \sum_{p=0}^{\infty}\!'\int_{1}^{\infty}du  \int_{ \xi_p d}^{\infty}dx\,x\ln\left(1- e^{-2ux}\right).
\end{equation*}The integral on the right hand side can be integrated explicitly which gives
\begin{equation}\label{eq11_25_2}
E_{\text{Cas}}^{\text{PFA}}\sim -\frac{   r_Ar_B T}{4d(r_B-r_A)}\sum_{p=0}^{\infty}\!'\sum_{k=1}^{\infty}\frac{1}{ k^3}e^{-4\pi kp dT}.
\end{equation}
By summing over $p$, we have
\begin{equation}\label{eq11_25_3} \begin{split}
E_{\text{Cas}}^{\text{PFA}}\sim  &-\frac{  r_Ar_B T}{8d(r_B-r_A)}\sum_{k=1}^{\infty}
\frac{\coth 2\pi kd T}{k^3}\\=&-\frac{  \pi^3r_Ar_B }{1440d^2(r_B-r_A)}\left(1+h_s(2dT)\right),
\end{split}\end{equation}where the function $h_s(x)$ is  given by
\begin{equation*}
h_s(x)=90x^4\sum_{k=1}^{\infty}\left(\frac{\coth \pi kx}{(\pi kx)^3}-\frac{1}{(\pi k x)^4}\right).
\end{equation*}Differentiating with respect to $d$ gives the proximity force approximation for the Casimir force:
\begin{equation}\label{eq11_25_4}
F_{\text{Cas}}^{\text{PFA}}\sim   -\frac{  \pi^3r_Ar_B }{720d^3(r_B-r_A)}\left(1+g_s(2dT)\right),
 \end{equation}where
\begin{equation*}
g_s(x)=h_s(x)-\frac{x}{2}h_s'(x)= 45x^4 \sum_{k=1}^{\infty}\left(\frac{1}{(\pi kx)^2}\frac{1}{ \sinh^2( \pi k x)}+\frac{\coth ( \pi k x)}{ (\pi k x)^3}\right)-1.
\end{equation*} The expressions \eqref{eq11_25_3} and \eqref{eq11_25_4} do not give us explicitly the leading terms of the proximity force approximations to the Casimir free energy and Casimir force. To find the leading behaviors, we need to return to \eqref{eq11_25_2}. From this expression, it is obvious that in the high temperature region where $dT\gg 1$, the dominating term is the sum of the terms with $p=0$, which gives
\begin{equation*}
E_{\text{Cas}}^{\text{PFA}}\sim -\frac{   r_Ar_B T}{8d(r_B-r_A)} \sum_{k=1}^{\infty}\frac{1}{ k^3}=-\frac{   r_Ar_B T\zeta_R(3)}{8d(r_B-r_A)},
\end{equation*}where $\zeta_R(s)=\sum_{n=1}^{\infty}n^{-s}$ is the Riemann zeta function. It follows that the proximity force approximation for the Casimir force is
\begin{equation*}
F_{\text{Cas}}^{\text{PFA}}\sim -\frac{   r_Ar_B T\zeta_R(3)}{8d^2(r_B-r_A)}.
\end{equation*}For the low temperature and medium temperature regions where $dT\ll 1$, we can use the inverse Mellin transform formula
\begin{equation*}
e^{-u}=\frac{1}{2\pi i}\int_{c-i\infty}^{c+i\infty} dz\Gamma(z) u^{-z}
\end{equation*} to transform \eqref{eq11_25_2} into
\begin{equation}\label{eq11_25_7}
E_{\text{Cas}}^{\text{PFA}}\sim -\frac{   r_Ar_B T\zeta_R(3)}{8d(r_B-r_A)}-\frac{   r_Ar_B T}{4d(r_B-r_A)}\frac{1}{2\pi i}\int_{c-i\infty}^{c+i\infty}
dz\Gamma(z)\zeta_R(z)\zeta_R(z+3)(4\pi dT)^{-z}.
\end{equation}Now the asymptotic behavior of $E_{\text{Cas}}^{\text{PFA}}$ when $dT\ll 1$ can be read off from the residues of the right hand side. Using
$$\Gamma(z)=\frac{2^{z-1}}{\sqrt{\pi}}\Gamma\left(\frac{z}{2}\right)\Gamma\left(\frac{z+1}{2}\right),$$we have
$$\Gamma(z)\zeta_R(z)\zeta_R(z+3)=\frac{2^{z}}{\sqrt{\pi}(z+1)}\Gamma\left(\frac{z}{2}\right)\zeta_R(z)\Gamma\left(\frac{z+3}{2}\right)\zeta_R(z+3).$$
Since $\Gamma(z/2)\zeta_R(z)$ only has poles at $z=0$ and $z=1$, we find that $\Gamma(z)\zeta_R(z)\zeta_R(z+3)$ only has poles at $z=-3,-2,-1,0,1$.
Evaluating the residues, we find that as $dT\ll 1$,
\begin{equation}\label{eq11_25_7}
E_{\text{Cas}}^{\text{PFA}}\sim  -\frac{   \pi^3r_Ar_B  }{1440d^2(r_B-r_A)} -\frac{\pi^3r_Ar_BT^2}{72(r_B-r_A)}+\frac{ r_Ar_BdT^3}{2 (r_B-r_A)} \zeta_R(3)-\frac{\pi^3r_Ar_B d^2T^4}{90(r_B-r_A)}.
\end{equation}  These are the only terms that are of polynomial order in $dT$. The remaining terms are exponentially small terms in $dT$. Differentiating with respect to $d$ gives the proximity force approximation to the Casimir force:
\begin{equation}\label{eq11_25_8}
F_{\text{Cas}}^{\text{PFA}}\sim  -\frac{   \pi^3r_Ar_B  }{720d^3(r_B-r_A)} -\frac{ r_Ar_BT^3}{2 (r_B-r_A)} \zeta_R(3)+\frac{\pi^3r_Ar_B dT^4}{45(r_B-r_A)}.
\end{equation}\eqref{eq11_25_7} and \eqref{eq11_25_8} give the leadings terms in the medium temperature region. The first terms on the right hand sides are the zero temperature leading terms, and the second terms are the leading terms for the temperature corrections. Notice that these leading terms of the temperature corrections are finite when $d\rightarrow 0$. In the low temperature region where $r_AT <r_BT\ll 1$, only the zero temperature leading terms are dominating.

In the case of RD or DR boundary conditions, the
proximity force approximation of the Casimir free energy gives the  leading behavior:
\begin{equation*}
E_{\text{Cas}}^{\text{PFA}}\sim \frac{  r_A r_BT}{d(r_B-r_A) }  \sum_{p=0}^{\infty}\!'\int_{1}^{\infty}du  \int_{d\xi_p}^{\infty}dx\,x\ln\left(1+ e^{-2ux}\right).
\end{equation*}Computing in exactly the same way as  in the case of DD or RR boundary conditions, we find that
\begin{equation}\label{eq11_25_3_2} \begin{split}
E_{\text{Cas}}^{\text{PFA}}\sim  &-\frac{  r_Ar_B T}{8d(r_B-r_A)}\sum_{k=1}^{\infty}(-1)^k
\frac{\coth 2\pi kd T}{k^3}\\=& \frac{ 7 \pi^3r_Ar_B }{11520d^2(r_B-r_A)}\left(1+h_a(2dT)\right),
\end{split}\end{equation}
\begin{equation}\label{eq11_25_4_2}
F_{\text{Cas}}^{\text{PFA}}\sim    \frac{  7\pi^3r_Ar_B }{5760d^3(r_B-r_A)}\left(1+g_a(2dT)\right),
 \end{equation}where the functions $h_a(x)$ and $g_a(x)$ are  given respectively by
\begin{equation*}\begin{split}
h_a(x)=&\frac{720x^4}{7}\sum_{k=1}^{\infty}(-1)^{k-1}\left(\frac{\coth \pi kx}{(\pi kx)^3}-\frac{1}{(\pi k x)^4}\right),\\
g_a(x)=&h_a(x)-\frac{x}{2}h_a'(x)= \frac{360}{7}x^4 \sum_{k=1}^{\infty}(-1)^{k-1}\left(\frac{1}{(\pi kx)^2}\frac{1}{ \sinh^2( \pi k x)}+\frac{\coth ( \pi k x)}{ (\pi k x)^3}\right)-1.
\end{split}\end{equation*}
In the high temperature region where $dT\gg 1$, the leading terms for the Casimir free energy and the Casimir force are respectively
\begin{equation*}
E_{\text{Cas}}^{\text{PFA}}\sim   \frac{ 3  r_Ar_B T\zeta_R(3)}{32d(r_B-r_A)},
\end{equation*}
\begin{equation*}
F_{\text{Cas}}^{\text{PFA}}\sim   \frac{ 3  r_Ar_B T\zeta_R(3)}{32d^2(r_B-r_A)}.
\end{equation*}When $dT\ll 1$,
\begin{equation}\label{eq11_25_9}
E_{\text{Cas}}^{\text{PFA}}\sim   \frac{  7 \pi^3r_Ar_B  }{11520d^2(r_B-r_A)} +\frac{\pi^3r_Ar_BT^2}{144(r_B-r_A)}  -\frac{\pi^3r_Ar_B d^2T^4}{90(r_B-r_A)},
\end{equation}
\begin{equation}\label{eq11_25_10}
F_{\text{Cas}}^{\text{PFA}}\sim   \frac{  7 \pi^3r_Ar_B  }{5760d^3(r_B-r_A)}  +\frac{\pi^3r_Ar_B dT^4}{45(r_B-r_A)}.
\end{equation}
As before, \eqref{eq11_25_9} and \eqref{eq11_25_10} give the leading terms in the medium temperature region. The first terms on the right hand sides   give the zero temperature leading terms, and the second terms are the leading terms of the temperature corrections. Notice that for the Casimir force, the leading term of the temperature correction is linear in $d$. The constant term is missing. In the low temperature region, the leading terms are the zero temperature leading terms.

\begin{figure}[h]\centering
\epsfxsize=0.5\linewidth \epsffile{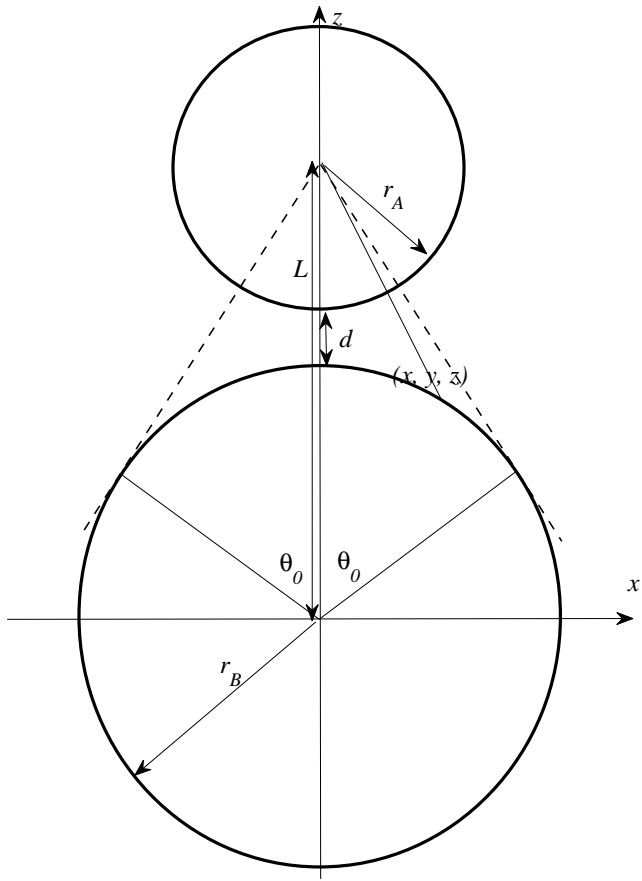} \caption{\label{f2}The cross section of   two spheres outside each other.}   \end{figure}

For the case where the two spheres are outside each other (see Fig. \ref{f2}), we only integrate over the part of the sphere $B$ with
$$0\leq \theta\leq \theta_0=\cos^{-1}\frac{r_B}{L}.$$ This gives
 \begin{equation*}
E_{\text{Cas}}^{\text{PFA}}= 2\pi r_B^2\int_0^{\theta_0}d\theta \sin\theta\mathcal{E}_{\text{Cas}}^{\parallel}(h(\theta)),
\end{equation*} where $h(\theta)$ is still given by \eqref{eq11_24_3} but $L=r_A+r_B+d$. Making a change of variable as before,
we have
\begin{equation*}
E_{\text{Cas}}^{\text{PFA}}=\frac{2\pi  r_B}{L}\int_{d}^{\sqrt{L^2-r_B^2}-r_A}du (u+r_A) \mathcal{E}_{\text{Cas}}^{\parallel}(u).
\end{equation*}The rest of the computations are the same as for the case where one sphere is inside the other.  The results are almost identical to the results above, except that one has to change the factor $(r_B-r_A)$ in the denominators to $(r_A+r_B)$.
In fact one finds that one would obtain the same results even if the integration is carried out  over the whole sphere $B$.

Finally, for electromagnetic field, it was well known that the Casimir free energy density for a pair of parallel   plates, where both are perfectly conducting or both are infinitely permeable, is twice of the Casimir free energy density for a pair of parallel Dircichlet plates. The Casimir free energy density for a pair of parallel plates, where one is perfectly conducting and one is infinitely permeable, is twice of the Casimir free energy density for a pair of parallel plates, where one is Dirichlet and one is Neumann. Therefore, for two spheres which are both perfectly conducting or both infinitely permeable, the proximity force approximations for the Casimir free energy and Casimir force are twice of that for two spheres which are both imposed with Dirichlet boundary conditions. For two spheres where one is perfectly conducting and one is infinitely permeable, the proximity force approximations for the Casimir free energy and Casimir force are twice of that for two spheres where one is imposed with Dirichlet boundary conditions and one is imposed with Robin boundary conditions. Namely,
\begin{equation*}
\begin{split}
E_{\text{Cas}}^{\text{PFA},\text{CC}}=&E_{\text{Cas}}^{\text{PFA},\text{PP}}=2E_{\text{Cas}}^{\text{PFA},\text{DD}},\\
E_{\text{Cas}}^{\text{PFA},\text{CP}}=&E_{\text{Cas}}^{\text{PFA},\text{PC}}=2E_{\text{Cas}}^{\text{PFA},\text{DR}}.
\end{split}
\end{equation*}

\section{Asymptotic  behaviors of the zero temperature Casimir energy and Casimir force from exact formulas}
In this section, we are going to compute the small distance asymptotic behavior of the zero temperature Casimir   energy   from the exact formulas given in Section \ref{s2}. The method we are using is similar to that used in \cite{1} to find the asymptotic behavior between a cylinder and a plate and in \cite{16,17,22} to find the asymptotic behavior between a sphere and a plate. In \cite{18}, we have extended this method to the case of two cylinders.   For two spheres, things are more involved because the $M$ matrix is more complicated. We will explain our method in detail for the case where sphere $A$ is inside sphere $B$. At the end of this section, we will discuss the necessary changes for the case where the two spheres are outside each other.

As we mentioned in Section \ref{s3}, we only consider the case where $L>0$, i.e., the spheres are eccentric. Let
$$\vep=\frac{d}{r_B-r_A},\hspace{1cm}a=\frac{r_A}{r_B-r_A},\hspace{1cm}b=\frac{r_B}{r_B-r_A}.$$ Obviously   $b=1+a$. We want to find the asymptotic behavior of the Casimir energy when $\vep\ll 1$. Making a change of variables $\xi=\omega/(r_B-r_A)$ and expanding the logarithm, the zero temperature Casimir energy for scalar fields \eqref{eq11_29_1} can be written as
\begin{equation*}
E_{\text{Cas}}^{\text{XY},T=0}=-\frac{1}{2\pi(r_B-r_A)}\sum_{s=0}^{\infty}\frac{1}{s+1}\int_0^{\infty} d\omega\sum_{m=-\infty}^{\infty}\sum_{l=|m|}^{\infty}\left(\prod_{j=1}^s\sum_{l_j=|m|}^{\infty}\right) \left(\prod_{i=0}^s\tilde{M}^{\text{XY}}_{l_i,l_{i+1}}(\omega)\right),
\end{equation*}where
\begin{equation*}
\tilde{M}^{\text{XY}}_{l_i,l_{i+1}}=(-1)^{l_i-l_{i+1}}M^{\text{XY}}_{l_i,l_{i+1}}=T^{A,\text{X}}_{i}
\sum_{\tilde{l}_i=|m|}^{\infty}\tilde{U}_{i}^{AB}T_{i}^{B,\text{Y}}\tilde{U}^{BA}_{i},
\end{equation*}with
\begin{align*}
T^{A,\text{D}}_{i}=&\frac{ I_{l_{i}+1/2}(\omega a) }{ K_{l_i+1/2}(\omega a) },\hspace{1cm}T^{A,\text{R}}_{i}=\frac{u_A I_{l_{i}+1/2}(\omega a)+\omega a I_{l_{i}+1/2}'(\omega a)}{u_A K_{l_i+1/2}(\omega a)+\omega aK_{l_{i}+1/2}'(\omega a) },\\
T^{B,\text{D}}_{i}=&\frac{ K_{\tilde{l}_i+1/2}(\omega b) }{ I_{\tilde{l}_i+1/2}(\omega b) },\hspace{1cm}T^{B,\text{R}}_{i}=\frac{ u_BK_{\tilde{l}_i+1/2}(\omega b)
 +\omega bK_{\tilde{l}_i+1/2}'(\omega b)}{ u_BI_{\tilde{l}_i+1/2}(\omega b)+\omega bI_{\tilde{l}_i+1/2}'(\omega b) };
\end{align*}\begin{equation*}\begin{split}
\tilde{U}_{i}^{ AB  }=(-1)^m \sqrt{\frac{\pi}{2\omega (1-\vep)}}\sum_{l_i^{\prime\prime}=|l_i-\tilde{l}_i|}^{l_i+\tilde{l}_i}\sqrt{(2l_i+1)(2\tilde{l}_i+1)}(2l_i^{\prime\prime}+1)\begin{pmatrix} l_i & \tilde{l}_i & l_i^{\prime\prime}\\ 0 & 0 & 0\end{pmatrix}
\begin{pmatrix} l_i & \tilde{l}_i & l_i^{\prime\prime}\\ m & -m & 0\end{pmatrix}
I_{l_i^{\prime\prime}+1/2}(\omega (1-\vep)),\end{split}
\end{equation*}and $\tilde{U}_{i}^{ BA  }$ is obtained from $\tilde{U}_{i}^{ AB  }$ by replacing $l_i$ with $l_{i+1}$, and $l_i^{\prime\prime}$ with $\tilde{l}_i^{\prime\prime}$.  Here we have used the fact that   $$\begin{pmatrix} \tilde{l}_i  & l_{i+1} & \tilde{l}_i^{\prime\prime}\\ m &-m &0\end{pmatrix}\begin{pmatrix} \tilde{l}_i  & l_{i+1} & \tilde{l}_i^{\prime\prime}\\ 0 &0&0\end{pmatrix}=(-1)^{l_{i+1}+\tilde{l}_i+\tilde{l}_i^{\prime\prime}}\begin{pmatrix} l_{i+1}  & \tilde{l}_{i} & \tilde{l}_i^{\prime\prime}\\ m &-m &0\end{pmatrix}\begin{pmatrix} l_{i+1}  & \tilde{l}_{i}  & \tilde{l}_i^{\prime\prime}\\ 0 &0&0\end{pmatrix},$$
which follows from the properties of the $3j$-symbols.
In the following, we also have to use the fact that the $3j$-symbols
\begin{equation*}
\begin{pmatrix} l_i  & \tilde{l}_{i} & l_i^{\prime\prime}\\ 0 &0&0\end{pmatrix}\quad\text{and}\quad\begin{pmatrix} \tilde{l}_i  & l_{i+1} & \tilde{l}_i^{\prime\prime}\\ 0 &0&0\end{pmatrix}\end{equation*}are nonzero if and only if $l_i+\tilde{l}_i+l_i^{\prime\prime}$ and $\tilde{l}_i+l_{i+1}+\tilde{l}_i^{\prime\prime}$ are even.

The small distance leading asymptotic behavior of the Casimir energy   comes from the regions where $l_i$ and $\omega$ are large.
Introduce new variables $n_1,\ldots, n_s$, $q_0,\ldots,q_s$, $\nu_0,\ldots,\nu_s$, $\tilde{\nu}_0,\ldots,\tilde{\nu}_s$ and $\tau$ such that
\begin{equation}\label{eq12_01_9}\begin{split}
l_i=&l+n_i,\quad 1\leq i\leq s,\\
\tilde{l}_i=&\frac{b}{2a}(l_i+l_{i+1})+q_i=\frac{b}{a}l+\frac{b}{2a}( n_i+n_{i+1})+q_i,\quad 0\leq i\leq s,\\
l_i^{\prime\prime}=&\tilde{l}_i-l_i+2\nu_i=\frac{l}{a}-n_i+\frac{b}{2a}(n_i+n_{i+1})+q_i+2\nu_i,\quad 0\leq i\leq s,\\
\tilde{l}_i^{\prime\prime}=& \tilde{l}_i-l_{i+1}+2\tilde{\nu}_i=\frac{l}{a}-n_{i+1}+\frac{b}{2a}(n_i+n_{i+1})+q_i+2\tilde{\nu}_i,\quad 0\leq i\leq s,\\
\omega=&\frac{l\sqrt{1-\tau^2}}{a\tau},
\end{split}
\end{equation}where $l=l_0$.
In terms of the new variables, the leading contribution to the Casimir energy comes from $l\sim \vep^{-1}$, $n_i,q_i,m\sim \vep^{-1/2}$ and $\nu_i,\tilde{\nu}_i,\tau \sim 1$. The summation over $l_i, 1\leq i\leq s$, is transformed to summation over $n_i$ from $n_i=|m|-l$ to $\infty$, which in the $\vep\ll 1$ limit can be replaced by an integration over $n_i$ from $-\infty$ to $\infty$. The summation over $\tilde{l}_i$ is transformed to summation over $q_i$ in the set $$\left\{|m|-\frac{b}{a}l-\frac{b}{2a}( n_i+n_{i+1})+n\,:\,n=0,1,2,\ldots\right\},$$which in the $\vep\ll 1$ limit can be replaced by integration over $q_i$ from $-\infty$ to $\infty$. For the summation over $l_i^{\prime\prime}$,  the condition $|l_i-\tilde{l}_i|\leq l_i^{\prime\prime}\leq l_i+\tilde{l}_i$ and the condition that $l_i+\tilde{l}_i+l_i^{\prime\prime}$ has to be even are equivalent to $\nu_i$ is an integer satisfying $\max\{0,l_i-\tilde{l}_i\}\leq \nu_i\leq l_i$. Therefore,  in the $\vep\ll 1$ limit, the summation over $l_i^{\prime\prime}$ can be replaced by the summation over $\nu_i$ from $0$ to $\infty$. Similarly, the summation over $\tilde{l}_i^{\prime\prime}$ can be replaced by the summation over $\tilde{\nu}_i$ from $0$ to $\infty$. Since
\begin{equation*}
\sum_{m=-\infty}^{\infty} \sum_{l=|m|}^{\infty} =\sum_{l=0}^{\infty}\sum_{m=-l}^{l},
\end{equation*}in the $\vep\ll 1$ limit, the summations over $l$ and $m$ can be replaced by integrations over $l$ and $m$ from $0$ to $\infty$ and from $-\infty$ to $\infty$ respectively. Hence,
\begin{equation}\label{eq11_30_9}
E_{\text{Cas}}^{\text{XY},T=0}\sim -\frac{1}{2\pi r_A}\sum_{s=0}^{\infty}\frac{1}{s+1}\int_0^{1} \frac{d\tau}{\tau^2\sqrt{1-\tau^2}} \int_{0}^{\infty}dl\,l\int_{-\infty}^{\infty}dm\left(\prod_{i=1}^s\int_{-\infty}^{\infty}dn_i\right) \left(\prod_{i=0}^s\tilde{M}^{\text{XY}}_{l_i,l_{i+1}}\right),
\end{equation}
with
\begin{equation*}
\tilde{M}^{\text{XY}}_{l_i,l_{i+1}}\sim T^{A,\text{X}}_{i}
\int_{-\infty}^{\infty} dq_i\tilde{U}_{i}^{AB}T_{i}^{B,\text{Y}}\tilde{U}^{BA}_{i}.
\end{equation*}
In the following, we will compute the leading order term and the next-to-leading order term of the zero temperature Casimir energy. For this, we need to expand $\tilde{M}^{\text{XY}}_{l_i,l_{i+1}}$ up to terms of order $\vep$. First, consider the $3j$-symbol $$\begin{pmatrix} l_i & \tilde{l}_i & l_i^{\prime\prime}\\ m & -m & 0\end{pmatrix}.$$ Although the asymptotic expansion of this $3j$-symbol has been derived in \cite{16}, it cannot be directly applied here since in our case, $l_i^{\prime\prime}=\tilde{l}_i-l_i+2\nu_i$ instead of $l_i^{\prime\prime}=\tilde{l}_i+l_i-2\nu_i$. Nevertheless, this small problem is easy to overcome. Using the property of $3j$-symbols, we have
 \begin{equation*}
\begin{pmatrix} l_i & \tilde{l}_i & l_i^{\prime\prime}\\ m & -m & 0\end{pmatrix}=
\begin{pmatrix} l_i^{\prime\prime} & l_i & \tilde{l}_i\\ 0 & m & -m\end{pmatrix}.
\end{equation*} Similar to the case considered in \cite{16}, we have the integral representation
  \begin{equation}\label{eq11_29_3}\begin{split}
\begin{pmatrix} l_i^{\prime\prime} & l_i & \tilde{l}_i\\ 0 & m & -m\end{pmatrix}=& (-1)^{ m }\frac{(-2i)^{l_i^{\prime\prime}+l_i+\tilde{l}_i}}{\pi^2}C_{l_i^{\prime\prime}0l_im}^{\tilde{l}_i m}
\int_{-\frac{\pi}{2}}^{\frac{\pi}{2}}d\theta \int_{-\frac{\pi}{2}}^{\frac{\pi}{2}}d\phi\,
e^{ 2im\theta}\cos^{l_i+\tilde{l}_i-l_i^{\prime\prime}}\theta\sin^{l_i+l^{\prime\prime}_i-\tilde{l}_i}(\theta-\phi)\cos^{l_i^{\prime\prime}+\tilde{l}_i-l_i}\phi \end{split}
\end{equation}from the work \cite{19}, where
\begin{equation*}
C_{l_i^{\prime\prime}0l_im}^{\tilde{l}_i m}=\left[\frac{ (l_i+m)! \,(l_i-m)!\,(\tilde{l}_i-m)!\,(\tilde{l}_i+m)!\,l_i^{\prime\prime}!\,l_i^{\prime\prime}!}
{(l_i+\tilde{l}_i+l_i^{\prime\prime}+1)!\,(l_i+\tilde{l}_i-l_i^{\prime\prime})!\,(l_i-\tilde{l}_i+l_i^{\prime\prime})!\,(-l_i+\tilde{l}_i+l_i^{\prime\prime})!}\right]^{\frac{1}{2}}.
\end{equation*}Then we can proceed as in \cite{16}.
For the factor in front of the integral, we can use Stirling's formula
$$\ln n!=\left(n+\frac{1}{2}\right)\ln n-n+\frac{1}{2}\ln 2\pi+\frac{1}{12 n}+\ldots$$
to obtain an expansion of the form
\begin{equation}\label{eq11_30_1}
 \frac{2^{l_i+\tilde{l}_i+l_i^{\prime\prime}}}{\pi^2}C_{l_i^{\prime\prime}0l_im}^{\tilde{l}_im}\sim\frac{2^{ \nu_i -\frac{1}{2}  }}{ \pi^{\frac{5}{4}}\sqrt{(2\nu_i)!}}\left(\frac{l}{b}\right)^{\nu_i+\frac{1}{4}} \exp\left(\frac{m^2(a+b)}{2bl}\right)\exp
\Bigl(\mathcal{A}_{i,1}+\mathcal{A}_{i,2} \Bigr),
\end{equation}where here and in the following, for any $\mathcal{X}$, $\mathcal{X}_{i,1}$ and $\mathcal{X}_{i,2}$ are respectively terms of order $\sqrt{\vep}$  and $\vep$.  All the expansions are Taylor expansions which can be done using a machine and we will not write down explicitly the expressions for  $\mathcal{X}_{i,1}$ and $\mathcal{X}_{i,2}$.
For the integral
  $$\mathfrak{J}=\int_{-\frac{\pi}{2}}^{\frac{\pi}{2}}d\theta \int_{-\frac{\pi}{2}}^{\frac{\pi}{2}}d\phi\,
e^{ 2im\theta}\cos^{l_i+\tilde{l}_i-l_i^{\prime\prime}}\theta\sin^{l_i+l^{\prime\prime}_i-\tilde{l}_i}(\theta-\phi)\cos^{l_i^{\prime\prime}+\tilde{l}_i-l_i}\phi,$$
use the fact that $\sin u=u(1-u^2/6+\ldots)$, $\cos u=\exp(-\ln\sec u)$ and $\ln\sec u=u^2/2+u^4/12+\ldots$.  Introduce new  variables $\rho$ and $\sigma$ of order $\vep^0$ so that
\begin{equation*}
\theta=\frac{\rho+\sigma}{\sqrt{ l }},\hspace{1cm} \phi=\frac{\rho-a\sigma}{\sqrt{ l }},
\end{equation*}one obtains an asymptotic expansion of the form
\begin{equation*}\begin{split}
\mathfrak{J}\sim &\frac{b^{2\nu_i+1}}{l^{\nu_i+1} }\int_{-\infty}^{\infty}d\sigma\int_{-\infty}^{\infty}d\rho\;\sigma^{2\nu_i}\left(1-\frac{b^2\sigma^2}{6 l}\right)^{2\nu_i}
\exp\left(\frac{2 im (\rho+\sigma)}{ \sqrt{ l}} -\frac{b}{a}\rho^2 - b\sigma^2 \right) \Bigl(1+\mathcal{B}_{i,1}+\mathcal{B}_{i,2}\Bigr).
\end{split}\end{equation*}
Making a change of variables $\tilde{\rho}=\rho-ima/(b\sqrt{l})$ and integrating with respect to $\tilde{\rho}$ with the help of a machine, one obtains
\begin{equation*}\mathfrak{J}
\sim  \frac{\sqrt{\pi  a }b^{2\nu_i+1/2}}{l^{\nu_i+1}}\exp\left(-\frac{am^2}{bl}\right)\int_{-\infty}^{\infty}d\sigma\, \sigma^{2\nu_i}
\exp\left(\frac{2 im \sigma}{ \sqrt{ l}} - b\sigma^2 \right)  \Bigl(1+\mathcal{C}_{i,1}+\mathcal{C}_{i,2}\Bigr) .\end{equation*}
Together with \eqref{eq11_30_1}, we have the expansion
\begin{equation}\label{eq11_30_3}\begin{split}
\begin{pmatrix} l_i & \tilde{l}_i & l^{\prime\prime}_i\\ m & -m & 0\end{pmatrix}  \sim &(-1)^m(-i)^{l_i^{\prime\prime}+l_i+\tilde{l}_i}\frac{2^{\nu_i -\frac{1}{2}  } \sqrt{a}b^{\nu_i+1/4}}{l^{\frac{3}{4}} \pi^{\frac{3}{4}}\sqrt{(2\nu_i)!} } \exp\left(\frac{m^2}{2bl}\right)\exp
\Bigl(\mathcal{A}_{i,1}+\mathcal{A}_{i,2}\Bigr)\\&\times \int_{-\infty}^{\infty}d\sigma \sigma^{2\nu_i}
\exp\left(\frac{2 im \sigma}{ \sqrt{ l}}  -b\sigma^2 \right)  \Bigl(1+\mathcal{C}_{i,1}+\mathcal{C}_{i,2}\Bigr).
\end{split}\end{equation}Setting $m=0$, one can integrate \eqref{eq11_30_3} with respect to $\sigma$ using
\begin{equation*}
\begin{split}
\int_{-\infty}^{\infty}d\sigma\;\sigma^{2\nu_i}e^{-b\sigma^2}=\frac{\Gamma\left(\nu_i+1/2\right)}{b^{\nu_i+1/2}}
=\frac{(2\nu_i)!}{2^{2\nu_i}\nu_i!}\frac{\sqrt{\pi}}{b^{\nu_i+1/2}}.
\end{split}
\end{equation*}These give an expansion of the form
\begin{equation}\label{eq11_30_4}\begin{split}
\begin{pmatrix} l_i & \tilde{l}_i & l^{\prime\prime}_i\\ 0 & 0 & 0\end{pmatrix}  \sim&(-i)^{l_i^{\prime\prime}+l_i+\tilde{l}_i}\frac{\sqrt{a}}{ b^{ \frac{1}{4}}}\frac{\sqrt{(2\nu_i)!}}{l^{\frac{3}{4}} \pi^{\frac{1}{4}}\nu_i!}2^{-\nu_i -\frac{1}{2}  }  \exp
\Bigl(\mathcal{D}_{i,1}+\mathcal{D}_{i,2}\Bigr)\Bigl(1+\mathcal{E}_{i,1}+\mathcal{E}_{i,2}\Bigr),
\end{split}\end{equation}where $\mathcal{D}_{i,j}=\mathcal{A}_{i,j}\Bigr|_{m=0}$ for $j=1,2$.
Finally for the term
\begin{equation*}
\mathfrak{T}=\sqrt{\frac{\pi}{2\omega (1-\vep)}}\sqrt{(2l_i+1)(2\tilde{l}_i+1)}(2l_i^{\prime\prime}+1) I_{l_i^{\prime\prime}+1/2}(\omega (1-\vep)),
\end{equation*}
  Debye asymptotic expansion of modified Bessel function gives
\begin{equation}\label{eq11_30_2}
\mathfrak{T}\sim \frac{ \sqrt{(2l_i+1)(2\tilde{l}_i+1)(2l_i^{\prime\prime}+1)}}{\sqrt{2\omega(1-\vep)}}
\exp\Bigl(\left[l_i^{\prime\prime}+1/2\right]\eta(\omega_1)\Bigr)(1+\omega_1^2)^{-\frac{1}{4}}\left(1+\frac{u_1(\omega_1)}{l_i^{\prime\prime}+1/2}+\ldots\right),
\end{equation}where
\begin{equation*}\begin{split}
\omega_1=&\frac{(1-\vep)\omega}{l_i^{\prime\prime}+1/2},\\ \eta(z)=&\sqrt{1+z^2}+\ln\frac{z}{1+\sqrt{1+z^2}},\\
u_1(z)=&\frac{1}{\sqrt{1+z^2}}\left(\frac{1}{8}-\frac{5}{24(1+z^2)}\right).\end{split}
\end{equation*}
Expanding up to terms of order $\vep$ gives an expansion of the form
\begin{equation}\label{eq11_30_5}\begin{split}
 \mathfrak{T}\sim &\frac{  2l\sqrt{b}\tau}{\sqrt{a}(1-\tau^2)^{1/4}}\left(\frac{1-\tau}{1+\tau}\right)^{\nu_i+\frac{l}{2a}+\frac{1}{4}+\frac{n_i+n_{i+1}}{4a}-\frac{n_i-n_{i+1}}{4}+\frac{q_i}{2}}
\\&\times\exp\left(\frac{l}{ a\tau}-\frac{\tau(n_i+n_{i+1}-an_i+an_{i+1})^2}{8al}-\frac{\tau q_i(n_i+n_{i+1}-an_i+an_{i+1})}{2l}-\frac{\tau a q_i^2}{2l}-\frac{\vep l}{a\tau}\right)\\&\times\Bigl(1+\mathcal{G}_{i,1}+\mathcal{G}_{i,2}\Bigr)\exp\Bigl(\mathcal{H}_{i,1}+\mathcal{H}_{i,2}\Bigr)\Bigl(1+\mathcal{J}_1\Bigr).
\end{split}\end{equation}Here the term $\Bigl(1+\mathcal{J}_1\Bigr)$ comes from the last bracket in \eqref{eq11_30_2}, with
$$\mathcal{J}_1=\frac{a\tau(3-5\tau^2)}{24l},$$which is of order $\vep$.
 Combining \eqref{eq11_30_3}, \eqref{eq11_30_4} and \eqref{eq11_30_5} gives
\begin{equation*}\begin{split}
\tilde{U}_i^{AB}\sim& \frac{ \tau\sqrt{ab}}{ \pi\sqrt{l}(1-\tau^2)^{1/4}}\left(\frac{1-\tau}{1+\tau}\right)^{ \frac{l}{2a}+\frac{1}{4}+\frac{n_i+n_{i+1}}{4a}-\frac{n_i-n_{i+1}}{4}+\frac{q_i}{2}}
\Bigl(1+\mathcal{J}_1\Bigr) \\&\times\exp\left(\frac{l}{ a\tau}-\frac{\tau(n_i+n_{i+1}-an_i+an_{i+1})^2}{8al}-\frac{\tau q_i(n_i+n_{i+1}-an_i+an_{i+1})}{2l}-\frac{\tau a q_i^2}{2l}-\frac{\vep l}{a\tau}+\frac{m^2}{2bl}\right)\\&\times\sum_{\nu_i=0}^{\infty}\left(\frac{1-\tau}{1+\tau}\right)^{\nu_i}\frac{b^{\nu_i }}{\nu_i!}\int_{-\infty}^{\infty} d\sigma\,\sigma^{2\nu_i}
\exp\left(\frac{2 im \sigma}{ \sqrt{ l}}  -b\sigma^2\right)  \Bigl(1+\mathcal{L}_{i,1}+\mathcal{L}_{i,2}\Bigr),
\end{split}\end{equation*}
where
\begin{equation*}\begin{split}
\mathcal{L}_{i,1}=&\mathcal{C}_{i,1}+\mathcal{E}_{i,1}+\mathcal{G}_{i,1}+\mathcal{K}_{i,1},\\
\mathcal{L}_{i,2}=&\mathcal{C}_{i,1}\mathcal{E}_{i,1}+ \mathcal{C}_{i,1}\mathcal{G}_{i,1}+\mathcal{C}_{i,1}\mathcal{K}_{i,1}+\mathcal{E}_{i,1}\mathcal{G}_{i,1}+\mathcal{E}_{i,1} \mathcal{K}_{i,1}+\mathcal{G}_{i,1}\mathcal{K}_{i,1}+\frac{1}{2}\mathcal{K}_{i,1}^2+\mathcal{C}_{i,2}+\mathcal{E}_{i,2} +\mathcal{G}_{i,2}+\mathcal{K}_{i,2},\\
\mathcal{K}_{i,j}=&\mathcal{A}_{i,j}+\mathcal{D}_{i,j}+\mathcal{H}_{i,j},\hspace{1cm} j=1,2.
\end{split}
\end{equation*}After summing over $\nu_i$   using
\begin{equation*}
\sum_{n=0}^{\infty}\frac{v^{n}}{n!}=e^{v},\hspace{1cm}\sum_{n=0}^{\infty}\frac{nv^{n}}{n!}=v e^{v},
\hspace{1cm}\sum_{n=0}^{\infty}n^2\frac{v^{n}}{n!}=(v^2+v)e^{v},
\end{equation*} make a change of variable $\tilde{\sigma}=\sigma-i(1+\tau)m/(2b\tau\sqrt{l})$ and integrate over $\tilde{\sigma}$. These give
\begin{equation*}\begin{split}
\tilde{U}_i^{AB}\sim& \frac{\sqrt{a \tau}}{ \sqrt{2 \pi l  }}\left(\frac{1-\tau}{1+\tau}\right)^{ \frac{l}{2a}+\frac{n_i+n_{i+1}}{4a}-\frac{n_i-n_{i+1}}{4}+\frac{q_i}{2}}
\Bigl(1+\mathcal{J}_1\Bigr) \Bigl(1+\mathcal{M}_{i,1}+\mathcal{M}_{i,2}\Bigr)\\&\times\exp\left(\frac{l}{ a\tau}-\frac{\tau(n_i+n_{i+1}-an_i+an_{i+1})^2}{8al}-\frac{\tau q_i(n_i+n_{i+1}-an_i+an_{i+1})}{2l}-\frac{\tau a q_i^2}{2l}-\frac{\vep l}{a\tau}-\frac{m^2}{2bl\tau}\right).\end{split}
\end{equation*}
Interchanging $n_i$ and $n_{i+1}$ gives the corresponding  expansion for $\tilde{U}_i^{BA}$.

Now consider $T_i^{B,\text{Y}}$. Debye asymptotic expansions of modified Bessel functions give
\begin{equation}\label{eq11_30_6}\begin{split}
T_i^{B,\text{D}}\sim &\pi e^{-2(\tilde{l}_i+1/2)\eta(\omega_2)}\left(1-\frac{2u_1(\omega_2)}{\tilde{l}_i+1/2}\right),\\
T_i^{B,\text{R}}\sim &-\pi e^{-2(\tilde{l}_i+1/2)\eta(\omega_2)}\left(1-\frac{2m_{1,u_B}(\omega_2)}{\tilde{l}_i+1/2}\right),
\end{split}\end{equation}where
\begin{equation*}\begin{split}
\omega_2=&\frac{\omega b}{\tilde{l}_i+1/2},\\
m_{1,c}(z)=&\frac{1}{\sqrt{1+z^2}}\left(c-\frac{3}{8}+\frac{7}{24(1+z^2)}\right),
\end{split}
\end{equation*}and $\eta(z)$ and $u_1(z)$ are as before. Expanding up to terms of order $\vep$ gives
\begin{equation*}\begin{split}
T_i^{B,\text{Y}}\sim &(-1)^y\pi\exp\left(-\frac{2l b}{a\tau }+\frac{aq_i^2\tau}{bl}+\frac{q_i\tau(n_i+n_{i+1})}{l}+\frac{b(n_i+n_{i+1})^2\tau}{4al}\right)\\&\times\left(\frac{1-\tau}{1+\tau}\right)^{ -\frac{lb}{a} -\frac{1}{2}-\frac{b(n_i+n_{i+1})}{2a} -q_i}\exp\Bigl(\mathcal{N}_{i,1}+\mathcal{N}_{i,2}\Bigr)\Bigl(1+\mathcal{J}_2^{\text{Y}}\Bigr),\end{split}
\end{equation*}where $y=0$ for Y $=$ D and $y=1$ for Y $=$ R. The term $\Bigl(1+\mathcal{J}_2^{\text{Y}}\Bigr)$ comes from the last bracket of the term $T_i^{B,\text{Y}}$ in \eqref{eq11_30_6}, with
$$\mathcal{J}_2^{\text{D}}=-\frac{a\tau(3-5\tau^2)}{12bl},\hspace{1cm}\mathcal{J}_2^{\text{R}}=-\frac{a\tau(24u_B-9+7\tau^2)}{12bl}.$$
Combining the expansions for $\tilde{U}_i^{AB}, \tilde{U}_i^{BA}$ and $T_i^{B,\text{Y}}$ gives an expansion of the form
\begin{equation}\label{eq11_30_8}\begin{split}
\int_{-\infty}^{\infty} dq_i\;\tilde{U}_i^{AB}T_i^{B,\text{Y}}\tilde{U}_i^{BA}\sim &(-1)^y\frac{a   \tau}{2 l }\exp\left(-\frac{2l }{\tau}+\frac{\tau (n_i+n_{i+1})^2}{4l}
-\frac{a\tau (n_i-n_{i+1})^2}{4l}-\frac{m^2}{bl\tau}-\frac{2\vep l}{ a\tau}\right)\left(\frac{1-\tau}{1+\tau}\right)^{-l-\frac{n_i+n_{i+1}}{2}-\frac{1}{2}},\\
&\times\Bigl(1+2\mathcal{J}_1+\mathcal{J}_2^{\text{Y}}\Bigr)\int_{-\infty}^{\infty}dq_i\exp\left(-\frac{\tau a^2q_i^2}{bl}\right) \Bigl(1+\mathcal{P}_{i,1}+\mathcal{P}_{i,2}\Bigr).
\end{split}\end{equation}Here
\begin{equation*}
\begin{split}
\mathcal{P}_{i,1}=&\mathcal{M}_{i,1}+\tilde{\mathcal{M}}_{i,1}+\mathcal{N}_{i,1},\\
\mathcal{P}_{i,2}=&\mathcal{M}_{i,1}\tilde{\mathcal{M}}_{i,1}+\mathcal{M}_{i,1}\mathcal{N}_{i,1}+\tilde{\mathcal{M}}_{i,1}\mathcal{N}_{i,1}
+\frac{1}{2}\mathcal{N}_{i,1}^2+\mathcal{M}_{i,2}+\tilde{\mathcal{M}}_{i,2}+\mathcal{N}_{i,2},\\
\tilde{\mathcal{M}}_{i,j}=&\mathcal{M}_{i,j}\Bigl(n_i\leftrightarrow n_{i+1}\Bigr),\hspace{1cm}j=1,2.
\end{split}
\end{equation*}
 The integration with respect to $q_i$ is straightforward. For $T_i^{A,\text{X}}$, Debye asymptotic expansions of modified Bessel functions give
\begin{equation} \label{eq11_30_7}\begin{split}
T_i^{A,\text{D}}\sim &\frac{1}{\pi}
\exp\left(2[l_i+1/2]\eta(\omega_3) \right) \left(1+\frac{2u_1(\omega_3)}{l_i+1/2} \right),\\
T_i^{A,\text{R}}\sim &-\frac{1}{\pi}
\exp\left(2[l_i+1/2]\eta(\omega_3) \right) \left(1+\frac{2m_{1,u_A}(\omega_3)}{l_i+1/2} \right),
\end{split}\end{equation}where
$$\omega_3=\frac{\omega a}{l_i+1/2},$$and $\eta(z), u_1(z)$ and $m_{1,c}(z)$ are as before. Expanding up to terms of order $\vep$ give
\begin{equation*}
T_i^{A,\text{X}}\sim \frac{(-1)^x}{\pi}
\left(\frac{1-\tau}{1+\tau}\right)^{l+n_i+\frac{1}{2}}\exp\left(\frac{2l}{\tau}-\frac{\tau n_i^2}{l}\right)\exp\Bigl(\mathcal{Q}_{i,1}+\mathcal{Q}_{i,2} \Bigr) \Bigl(1+\mathcal{J}_3^{\text{X}}\Bigr),
\end{equation*}where $x=0$ if X $=$ D, and $x=1$ if X $=$ R. The term $\Bigl(1+\mathcal{J}_3^{\text{X}}\Bigr)$ comes from the last bracket of the term $T_i^{A,\text{X}}$ in \eqref{eq11_30_7}, with
$$\mathcal{J}_3^{\text{D}}=\frac{\tau(3-5\tau^2)}{12l},\hspace{1cm}\mathcal{J}_3^{\text{R}}=\frac{\tau(24u_A-9+7\tau^2)}{12l}.$$
Combining with \eqref{eq11_30_8}, we have
\begin{equation}\label{eq11_30_10}\begin{split}
\tilde{M}_{l_i,l_{i+1}}^{\text{XY}}\sim &(-1)^{x+y}\exp\left(-\frac{\tau (n_i^2-n_{i+1}^2)}{2l}\right)\left(\frac{1-\tau}{1+\tau}\right)^{ \frac{n_i-n_{i+1}}{2}}\frac{\sqrt{ b\tau  }}{2 \sqrt{\pi l }}\exp \left(-\frac{b\tau}{4l}(n_i-n_{i+1})^2-\frac{m^2}{bl\tau}-\frac{2\vep l}{a\tau}\right)\\&\Bigl(1+\mathcal{J}^{\text{XY}}\Bigr)\Bigl(1+\mathcal{S}_{i,1}+\mathcal{S}_{i,2}\Bigr),\end{split}
\end{equation}where
\begin{equation*}\begin{split}
\mathcal{S}_{i,1}=&\mathcal{R}_{i,1}+\mathcal{Q}_{i,1},\\
\mathcal{S}_{i,2}=&\mathcal{R}_{i,1}\mathcal{Q}_{i,1}+\frac{\mathcal{Q}_{i,1}^2}{2}+\mathcal{R}_{i,2}+\mathcal{Q}_{i,2},\\
\mathcal{R}_{i,j}=&\frac{a\sqrt{\tau}}{\sqrt{\pi bl}}\int_{-\infty}^{\infty}dq_i\exp\left(-\frac{\tau a^2q_i^2}{bl}\right)\mathcal{P}_{i,j},\quad j=1,2,\\
\mathcal{J}^{\text{XY}}=& 2\mathcal{J}_1+\mathcal{J}_2^{\text{Y}}+\mathcal{J}_3^{\text{X}}.
\end{split}\end{equation*}
Substitute \eqref{eq11_30_10} into \eqref{eq11_30_9}, we can compute the leading order term and the next-to-leading order term of the zero temperature Casimir energy:
\begin{equation*}\begin{split}
E^{\text{XY},T=0}_{\text{Cas}}\sim &-\frac{1}{2\pi r_A}\sum_{s=0}^{\infty}(-1)^{(x+y)(s+1)}\frac{b^{\frac{s+1}{2}}}{2^{s+1}\pi^{\frac{s+1}{2}}(s+1)}\int_0^{1} d\tau\frac{\tau^{\frac{s-3}{2}}}{ \sqrt{1-\tau^2}} \int_{0}^{\infty}dl\,l^{-\frac{s-1}{2}}\int_{-\infty}^{\infty}dm\left(\prod_{i=1}^s\int_{-\infty}^{\infty}dn_i\right) \\&
\times \exp \left(-\frac{b\tau}{4l}\sum_{i=0}^{s+1}(n_i-n_{i+1})^2-\frac{m^2(s+1)}{bl\tau}-\frac{2\vep l(s+1)}{a\tau}\right) \left(1+\sum_{i=0}^{s-1}\sum_{j=i+1}^{s}\mathcal{S}_{i,1}\mathcal{S}_{j,1}+\sum_{i=0}^s\mathcal{S}_{i,2}+(s+1)\mathcal{J}^{\text{XY}}\right).\end{split}\end{equation*}
We purposely separate the contribution from $\mathcal{J}^{\text{XY}}$ because besides the factor $(-1)^{(x+y)(s+1)}$, this is the only part that depends on the boundary conditions on the spheres. This term is independent of $n_i$ and $m$. The integration over $n_i$ can be performed as explained in \cite{1,18}, and then the integration over $m$ is also straightforward. After these, we obtain an expression of the form
\begin{equation}\label{eq11_30_11}\begin{split}
E^{\text{XY}, T=0}_{\text{Cas}}\sim &
 -\frac{b}{4\pi r_A}\sum_{s=0}^{\infty}\frac{(-1)^{(x+y)(s+1)}}{ (s+1)^{2}}\int_0^{1} \frac{\tau^{-1}d\tau}{ \sqrt{1-\tau^2}}\int_{0}^{\infty}dl\,l \exp \left( -\frac{2\vep l(s+1)}{a\tau}\right) \left(1+\mathcal{T}+(s+1)\mathcal{J}^{\text{XY}}\right).
\end{split}\end{equation}The integrations over $l$ and $\tau$ are also straightforward. For the DD case, we obtain
\begin{equation}\label{eq11_30_12}\begin{split}
E^{\text{DD}, T=0}_{\text{Cas}}\sim&-\frac{r_A r_B}{16\pi d^2(r_B-r_A)}\sum_{s=0}^{\infty}\frac{1}{ (s+1)^{4}}\left(1+\vep\left[\frac{1}{3ab}+1\right]\right)\\
=&-\frac{\pi^3 r_Ar_B}{1440 d^2(r_B-r_A)}\left(1+\frac{d}{r_B-r_A}+\frac{1}{3}\left[\frac{d}{r_A}-\frac{d}{r_B}\right]\right).
\end{split}\end{equation}
The other cases can be easily obtained from this. In the RD case, since
\begin{equation*}\mathcal{J}^{\text{RD}}-\mathcal{J}^{\text{DD}}=\mathcal{J}_3^{\text{R}}-\mathcal{J}_3^{\text{D}}=\frac{\tau(\tau^2-1+2u_A)}{l},\end{equation*}
we can infer from \eqref{eq11_30_11} and \eqref{eq11_30_12} that
\begin{equation*}
\begin{split}
E^{\text{RD}, T=0}\sim &-\frac{r_A r_B}{16\pi d^2(r_B-r_A)}\sum_{s=0}^{\infty}\frac{(-1)^{s+1}}{ (s+1)^{4}}\left(1+\vep\left[\frac{1}{3ab}+1\right]\right)\\
&-\frac{b}{4\pi r_A}\sum_{s=0}^{\infty}\frac{(-1)^{s+1}}{ (s+1)^{2}}\int_0^{1} \frac{\tau^{-1}d\tau}{ \sqrt{1-\tau^2}}\int_0^{\infty}dl\,l \exp \left( -\frac{2\vep l(s+1)}{a\tau}\right) \left((s+1)\frac{\tau(\tau^2-1+2u_A)}{l}\right)\\
=&\frac{7\pi^3 r_Ar_B}{11520 d^2(r_B-r_A)}\left(1+\frac{d}{r_B-r_A}+\frac{1}{3}\left[\frac{d}{r_A}-\frac{d}{r_B}\right]\right)
-\frac{b}{24\pi d}\sum_{s=0}^{\infty}\frac{(-1)^{s+1}}{(s+1)^2}(6u_A-1)\\
=&\frac{7\pi^3 r_Ar_B}{11520 d^2(r_B-r_A)}\left(1+\frac{d}{r_B-r_A}+\frac{1}{3}\left[\frac{d}{r_A}-\frac{d}{r_B}\right]+\frac{40}{7\pi^2}\frac{d}{r_A}(6u_A-1)\right).
\end{split}
\end{equation*}
In the DR case,
\begin{equation*}\mathcal{J}^{\text{RD}}-\mathcal{J}^{\text{DD}}=\mathcal{J}_2^{\text{R}}-\mathcal{J}_2^{\text{D}}=-\frac{a\tau(\tau^2-1+2u_B)}{bl}.\end{equation*}
Compare to the RD case, it is easy to see that
\begin{equation*}
E^{\text{DR}, T=0}\sim \frac{7\pi^3 r_Ar_B}{11520 d^2(r_B-r_A)}\left(1+\frac{d}{r_B-r_A}+\frac{1}{3}\left[\frac{d}{r_A}-\frac{d}{r_B}\right]-\frac{40}{7\pi^2}\frac{d}{r_B}(6u_B-1)\right).
\end{equation*}
Finally, for the RR case, we have
\begin{align*}
E^{\text{RR}, T=0}\sim &-\frac{r_A r_B}{16\pi d^2(r_B-r_A)}\sum_{s=0}^{\infty}\frac{1}{ (s+1)^{4}}\left(1+\vep\left[\frac{1}{3ab}+1\right]\right)
-\frac{b}{24\pi d}\sum_{s=0}^{\infty}\frac{1}{(s+1)^2}\left([6u_A-1]-\frac{a}{b}[6u_B-1]\right)\\=&-\frac{\pi^3 r_Ar_B}{1440 d^2(r_B-r_A)}\left(1+\frac{d}{r_B-r_A}+\frac{1}{3}\left[\frac{d}{r_A}-\frac{d}{r_B}\right]+\frac{10}{ \pi^2}\frac{d}{r_A}(6u_A-1)-\frac{10}{\pi^2}\frac{d}{r_B}(6u_B-1)\right).
\end{align*}

Next we consider the case of electromagnetic fields. From \eqref{eq12_01_1}, we have
\begin{equation}\label{eq12_01_6}
E_{\text{Cas}}^{\text{WZ},T=0}=-\frac{1}{2\pi(r_B-r_A)}\sum_{s=0}^{\infty}\frac{1}{s+1}\int_0^{\infty} d\omega\sum_{m=0}^{\infty}\sum_{l=\min\{1,|m|\}}^{\infty}\left(\prod_{j=1}^s\sum_{l_j=\min\{1,|m|\}}^{\infty}\right) \text{tr}\,\left(\prod_{i=0}^s\tilde{\mathbb{M}}^{\text{WZ}}_{l_i,l_{i+1}}(\omega)\right),
\end{equation}where
\begin{equation*}
\tilde{\mathbb{M}}^{\text{WZ}}_{l_i,l_{i+1}} =\mathbb{T}^{A,\text{W}}_{i}
\sum_{\tilde{l}_i=\min\{1,|m|\}}^{\infty}\tilde{\mathbb{U}}_{i}^{AB}\mathbb{T}_{i}^{B,\text{Z}}\tilde{\mathbb{U}}^{BA}_{i}.
\end{equation*}
Here\begin{equation*}
\tilde{\mathbb{U}}_{i}^{AB}=\tilde{U}_{i}^{AB}\begin{pmatrix}\Lambda_{l_i\tilde{l}_i}^{l_i^{\prime\prime}} & \tilde{\Lambda}_{l_i\tilde{l}_i}\\ \tilde{\Lambda}_{l_i\tilde{l}_i} &\Lambda_{l_i\tilde{l}_i}^{l_i^{\prime\prime}}\end{pmatrix},
\end{equation*}with
\begin{equation*}
\Lambda_{l_i\tilde{l}_i}^{l_i^{\prime\prime}}=\frac{1}{2}\frac{l_i^{\prime\prime}(l_i^{\prime\prime}+1)-l_i(l_i+1)-\tilde{l}_i(\tilde{l}_i+1)}
{\sqrt{l_i(l_i+1)\tilde{l}_i(\tilde{l}_i+1)}},\hspace{1cm}
 \tilde{\Lambda}_{l_i\tilde{l}_i}=\frac{ m\omega(1-\vep)}{\sqrt{l_i(l_i+1)\tilde{l}_i(\tilde{l}_i+1)}},
\end{equation*}$\tilde{\mathbb{U}}_i^{BA}$ is obtained from $\tilde{\mathbb{U}}_i^{AB}$ by interchanging $l_i$ and $l_{i+1}$;
 \begin{equation*}
\mathbb{T}_i^{*} =\begin{pmatrix} T_i^{*,\text{TE}}& 0\\ 0& T_i^{*,\text{TM}}  \end{pmatrix},
\end{equation*}where
\begin{equation}\label{eq12_01_3}
\begin{split}
T_i^{*,\text{C},\text{TE}}=&T_i^{*,\text{D}},\hspace{1cm}T_i^{*,\text{C},\text{TM}}=-T_i^{*,\text{R}}\Bigr|_{u_{*}=1/2},\\
T_i^{*,\text{P},\text{TE}}=&T_i^{*,\text{R}}\Bigr|_{u_{*}=1/2},\hspace{1cm}T_i^{*,\text{P},\text{TM}}=-T_i^{*,\text{D}}.
\end{split}
\end{equation}We need to expand $\Lambda_{l_i\tilde{l}_i}^{l_i^{\prime\prime}}$ up to terms of order $\vep$ and $\tilde{\Lambda}_{l_i\tilde{l}_i}$ up to terms of order $\sqrt{\vep}$, which gives
\begin{equation*}\begin{split}
\Lambda_{l_i\tilde{l}_i}^{l^{\prime\prime}_i}\sim&-\left(1-\frac{1}{2lb}-\frac{2\nu_i}{lb}\right)=-(1+\mathcal{U}_i),\\ \tilde{\Lambda}_{l\tilde{l};m}\sim&\frac{m\sqrt{1-\tau^2}}{bl\tau}=\mathcal{V}.\end{split}
\end{equation*}$\mathcal{U}_i$ is a term of order $\vep$ and $\mathcal{V}$ is a term of order $\sqrt{\vep}$.
Compare to the scalar case, it is easy to see that $\mathbb{M}^{\text{WZ}}_{l_i,l_{i+1}}$ has an expansion of the form:
\begin{equation}\label{eq12_01_5}\begin{split}
\mathbb{M}_{l_i,l_{i+1}}^{\text{WZ}}\sim&\exp\left(-\frac{\tau (n_i^2-n_{i+1}^2)}{2l}\right)\left(\frac{1-\tau}{1+\tau}\right)^{ \frac{n_i-n_{i+1}}{2}}\frac{\sqrt{ b\tau  }}{2 \sqrt{\pi l }}\exp \left(-\frac{b\tau}{4l}(n_i-n_{i+1})^2-\frac{m^2}{bl\tau}-\frac{2\vep l}{a\tau}\right)\Bigl(1+2\mathcal{J}_1\Bigr)\Bigl(1+\mathcal{S}_{i,1}+\mathcal{S}_{i,2}\Bigr)\\
&\times (-1)^{w+z}\begin{pmatrix} 1+\mathcal{J}_3^{\text{Z},\text{TE}} & 0 \\ 0 & 1+\mathcal{J}_3^{\text{Z},\text{TM}}\end{pmatrix}
\begin{pmatrix}-(1+\widehat{\mathcal{U}}) & \mathcal{V} \\ \mathcal{V}  & -(1+\widehat{\mathcal{U}})\end{pmatrix} \begin{pmatrix} 1+\mathcal{J}_2^{\text{W},\text{TE}} & 0 \\ 0 & 1+\mathcal{J}_2^{\text{W},\text{TM}} \end{pmatrix}\begin{pmatrix}-(1+\widehat{\mathcal{U}}) & \mathcal{V} \\ \mathcal{V}  & -(1+\widehat{\mathcal{U}})\end{pmatrix}.\end{split}\end{equation}
Here $w=0$ if W $=$ C and $w=1$ if W $=$ P, $z=0$ if Z $=$ C and $z=1$ if Z $=$ P,
\begin{equation*}
\widehat{\mathcal{U}}=\frac{\sqrt{2 b\tau}}{\sqrt{\pi(1+\tau)}}\exp\left(-\frac{m^2(1+\tau)}{2bl\tau}\right)\sum_{\nu_i=0}^{\infty}\left(\frac{1-\tau}{1+\tau}\right)^{\nu_i}\frac{b^{\nu_i}}{\nu_i!}
\int_{-\infty}^{\infty}d\sigma\,\sigma^{2\nu_i}\exp\left(\frac{2im\sigma}{\sqrt{l}}-b\sigma^2\right)
\mathcal{U}_i =-\frac{1}{2bl\tau}+\frac{m^2(1-\tau^2)}{2l^2b^2\tau^2}.
\end{equation*}
Using \eqref{eq12_01_3}, we find that for $k=2,3$,
\begin{equation*}\begin{split}
\mathcal{J}_k^{\text{C},\text{TE}}=\mathcal{J}_k^{\text{D}},\hspace{1cm}\mathcal{J}_k^{\text{C},\text{TM}}=\mathcal{J}_k^{\text{R} }\Bigr|_{u=1/2},\\
\mathcal{J}_k^{\text{P},\text{TE}}=\mathcal{J}_k^{\text{R}}\Bigr|_{u=1/2},\hspace{1cm}\mathcal{J}_k^{\text{P},\text{TM}}=\mathcal{J}_k^{\text{D} }.
\end{split}
\end{equation*}
Multiplying up the four matrices in \eqref{eq12_01_5} and keeping only terms up to order $\vep$ in the diagonal and terms up to order $\sqrt{\vep}$ in the off-diagonal, we have
\begin{equation}\label{eq12_01_2}\begin{split}
&\begin{pmatrix} 1+\mathcal{J}_3^{\text{Z},\text{TE}} & 0 \\ 0 & 1+\mathcal{J}_3^{\text{Z},\text{TM}}\end{pmatrix}
\begin{pmatrix}-(1+\widehat{\mathcal{U}}) & \mathcal{V} \\ \mathcal{V}  & -(1+\widehat{\mathcal{U}})\end{pmatrix} \begin{pmatrix} 1+\mathcal{J}_2^{\text{W},\text{TE}} & 0 \\ 0 & 1+\mathcal{J}_2^{\text{W},\text{TM}} \end{pmatrix}\begin{pmatrix}-(1+\widehat{\mathcal{U}}) & \mathcal{V} \\ \mathcal{V}  & -(1+\widehat{\mathcal{U}})\end{pmatrix}\\\sim &
\begin{pmatrix}1+\mathcal{J}_2^{\text{Z},\text{TE}}+\mathcal{J}_3^{\text{W},\text{TE}}+2\widehat{\mathcal{U}} +\mathcal{V}^2& -2\mathcal{V} \\ -2\mathcal{V}  & 1+\mathcal{J}_2^{\text{Z},\text{TM}}+\mathcal{J}_3^{\text{W},\text{TM}}+2\widehat{\mathcal{U}} +\mathcal{V}^2\end{pmatrix}.
\end{split}\end{equation}This term is independent of $i$. When we take the trace tr on $\prod_{i=0}^s\tilde{\mathbb{M}}^{\text{WZ}}_{l_i,l_{i+1}}$, we need to take the trace of the multiplication of $s+1$ copies of the matrix in \eqref{eq12_01_2}, which up to terms of order $\vep$, is given by
\begin{equation}\label{eq12_01_4}
\begin{split}
&\text{tr}\,\begin{pmatrix}1+\mathcal{J}_2^{\text{Z},\text{TE}}+\mathcal{J}_3^{\text{W},\text{TE}}+2\widehat{\mathcal{U}} +\mathcal{V}^2& -2\mathcal{V} \\ -2\mathcal{V}  & 1+\mathcal{J}_2^{\text{Z},\text{TM}}+\mathcal{J}_3^{\text{W},\text{TM}}+2\widehat{\mathcal{U}} +\mathcal{V}^2\end{pmatrix}^{s+1}\\
\sim & 2+(s+1)\left[\mathcal{J}_2^{\text{Z},\text{TE}}+\mathcal{J}_3^{\text{W},\text{TE}}\right]+(s+1)\left[\mathcal{J}_2^{\text{Z},\text{TM}}+\mathcal{J}_3^{\text{W},\text{TM}}\right]
+4(s+1)\widehat{\mathcal{U}}+2(s+1)\mathcal{V}^2+4s(s+1)\mathcal{V}^2.
\end{split}\end{equation}
Substituting \eqref{eq12_01_5} and \eqref{eq12_01_4} into \eqref{eq12_01_6} and compare to the scalar case, we find that
\begin{equation*}
E_{\text{Cas}}^{\text{WZ},T=0}\sim E_{\text{Cas}}^{\text{WZ},\text{sc},T=0} +\Delta E_{\text{Cas}}^{T=0},
\end{equation*}where
\begin{equation*}\begin{split}
\Delta E=&-\frac{1}{2\pi r_A}\sum_{s=0}^{\infty}\frac{(- 1)^{(w+z)(s+1)}b^{\frac{s+1}{2}}}{2^{s+1}\pi^{\frac{s+1}{2}}(s+1)}\int_0^{1} d\tau\frac{\tau^{\frac{s-3}{2}}}{ \sqrt{1-\tau^2}} \int_{0}^{\infty}dl\,l^{-\frac{s-1}{2}}\int_{-\infty}^{\infty}dm\left(\prod_{i=1}^s\int_{-\infty}^{\infty}dn_i\right) \\&
\times \exp \left(-\frac{b\tau}{4l}\sum_{i=0}^{s+1}(n_i-n_{i+1})^2-\frac{m^2(s+1)}{bl\tau}-\frac{2\vep l(s+1)}{a\tau}\right) \Bigl(4(s+1)\widehat{\mathcal{U}}+ 2(2s+1)(s+1)\mathcal{V}^2\Bigr),
\end{split}\end{equation*} 
\begin{equation*}\begin{split}
E_{\text{Cas}}^{ \text{CC},\text{sc},T=0} =E_{\text{Cas}}^{\text{DD},T=0}+E_{\text{Cas}}^{\text{RR},T=0}\Bigr|_{u_A=u_B=1/2}
\end{split}\end{equation*}up to the next-to-leading order term, and
\begin{equation*}\begin{split}
E_{\text{Cas}}^{ \text{CP},\text{sc},T=0} =E_{\text{Cas}}^{\text{RD},T=0}\Bigr|_{u_A=1/2}+E_{\text{Cas}}^{\text{DR},T=0}\Bigr|_{u_B =1/2}
\end{split}\end{equation*}up to the next-to-leading order term.
Since
\begin{equation*}
4(s+1)\widehat{\mathcal{U}}+2(2s+1)(s+1)\mathcal{V}^2=-\frac{2(s+1)}{bl\tau}+4(s+1)^2\frac{m^2(1-\tau^2)}{b^2l^2\tau^2},
\end{equation*}it is straightforward to find that
\begin{equation*}
\Delta E=\frac{1}{4\pi d}\sum_{s=0}^{\infty}\frac{(-1)^{(w+z)(s+1)}}{(s+1)^2}.
\end{equation*}
Combining with the results for the scalar case, we find that  
\begin{equation*}
E_{\text{Cas}}^{\text{CC},T=0}\sim  -\frac{\pi^3 r_Ar_B}{720 d^2(r_B-r_A)}\left(1+\frac{d}{r_B-r_A}+\left[\frac{1}{3}-\frac{20}{\pi^2}\right]\left[\frac{d}{r_A}-\frac{d}{r_B}\right]\right),
\end{equation*}
and
\begin{equation*}
E_{\text{Cas}}^{\text{CP},T=0}\sim \frac{7\pi^3 r_Ar_B}{5760 d^2(r_B-r_A)}\left(1+\frac{d}{r_B-r_A}+\left[\frac{1}{3}-\frac{80}{7\pi^2}\right]\left[\frac{d}{r_A}-\frac{d}{r_B}\right]\right).
\end{equation*}

In the following, we will briefly discuss the differences when the two spheres are outside each other. In this case, the small parameter is
$$\vep=\frac{d}{r_A+r_B},$$ and $\xi$ is related to $\omega$ by $\xi=\omega/(r_A+r_B)$. The   variables $\nu_i$ and $\tilde{\nu}_i$ are defined so that $l_i^{\prime\prime}=l_i+\tilde{l}_i-2\nu_i$ and
$\tilde{l}_i^{\prime\prime}=l_i+\tilde{l}_i-2\tilde{\nu}_i$. Then one has to use the integral representation given in \cite{16} for the $3j$-symbol. The other steps are the same. We find that the results are similar to the case where the sphere $A$ is inside the sphere $B$, one has only to change $r_B$ to $-r_B$. This can be understood as the change in the sign of the curvature of sphere $B$ when one change position from inside the sphere to outside the sphere. In the case of DD, NN and CC boundary conditions, our results agree with that obtained in \cite{20} using derivative expansion.

As a summary,   we have obtained the first two leading terms of the zero temperature Casimir energy in this section. For the scalar case, we have
\begin{equation}\label{eq12_15_1}
\begin{split}
E^{\text{DD}, T=0}_{\text{Cas}}\sim & -\frac{\pi^3 r_Ar_B}{1440 d^2(r_B\mp r_A)}\left(1\pm\frac{d}{r_B\mp r_A}+\frac{1}{3}\left[\frac{d}{r_A}\mp\frac{d}{r_B}\right]\right),\\
E^{\text{RR}, T=0}\sim &-\frac{\pi^3 r_Ar_B}{1440 d^2(r_B\mp r_A)}\left(1\pm \frac{d}{r_B\mp r_A}+\frac{1}{3}\left[\frac{d}{r_A}\mp \frac{d}{r_B}\right]+\frac{20}{ \pi^2}\frac{d}{r_A}(3\alpha_A-2)\mp \frac{20}{\pi^2}\frac{d}{r_B}(3\alpha_B-2)\right),\\
E^{\text{RD}, T=0}\sim & \frac{7\pi^3 r_Ar_B}{11520 d^2(r_B\mp r_A)}\left(1\pm \frac{d}{r_B\mp r_A}+\frac{1}{3}\left[\frac{d}{r_A}\mp\frac{d}{r_B}\right]+\frac{80}{7\pi^2}\frac{d}{r_A}(3\alpha_A-2)\right),\\
E^{\text{DR}, T=0}\sim &\frac{7\pi^3 r_A r_B}{11520 d^2(r_B\mp r_A)}\left(1\pm\frac{d}{r_B\mp r_A}+\frac{1}{3}\left[\frac{d}{r_A}\mp\frac{d}{r_B}\right]\mp\frac{80}{7\pi^2}\frac{d}{r_B}(3\alpha_B-2)\right),
\end{split}
\end{equation}
and for the electromagnetic case,
\begin{equation}\label{eq12_15_2}\begin{split}
E_{\text{Cas}}^{\text{CC},T=0} \sim  -\frac{\pi^3 r_Ar_B}{720 d^2(r_B\mp r_A)}\left(1\pm \frac{d}{r_B\mp r_A}+\left[\frac{1}{3}-\frac{20}{\pi^2}\right]\left[\frac{d}{r_A}\mp \frac{d}{r_B}\right]\right),\\
E_{\text{Cas}}^{\text{CP},T=0} \sim \frac{7\pi^3 r_Ar_B}{5760 d^2(r_B\mp r_A)}\left(1\pm \frac{d}{r_B\mp r_A}+\left[\frac{1}{3}-\frac{80}{7\pi^2}\right]\left[\frac{d}{r_A}\mp \frac{d}{r_B}\right]\right).
\end{split}\end{equation}
For the terms $\pm$ and $\mp$, the sign on the top is for the case where sphere $A$ is inside sphere $B$, and the sign in the bottom is for the case where the two spheres are exterior to each other. As one should expect, when the two spheres are outside each other, all these results are symmetric with respect to $r_A$ and $r_B$. Note that the leading terms agree with those obtained using proximity force approximation.

In the scalar case, specialize the Robin conditions to Neumann conditions by setting the Robin parameters $\alpha_A$ and $\alpha_B$ equal to zero, we have
\begin{equation*}
\begin{split}
E^{\text{NN}, T=0}\sim &-\frac{\pi^3 r_Ar_B}{1440 d^2(r_B\mp r_A)}\left(1\pm\frac{d}{r_B\mp r_A}+\left[\frac{1}{3}-\frac{40}{\pi^2}\right]\left[\frac{d}{r_A}\mp\frac{d}{r_B}\right] \right),\\
E^{\text{ND}, T=0}\sim & \frac{7\pi^3 r_Ar_B}{11520 d^2(r_B\mp r_A)}\left(1\pm \frac{d}{r_B\mp r_A}+\frac{1}{3}\left[\frac{d}{r_A}\mp\frac{d}{r_B}\right]-\frac{160}{7\pi^2}\frac{d}{r_A} \right),\\
E^{\text{DN}, T=0}\sim &\frac{7\pi^3 r_Ar_B}{11520 d^2(r_B\mp r_A)}\left(1\pm \frac{d}{r_B\mp r_A}+\frac{1}{3}\left[\frac{d}{r_A}\mp \frac{d}{r_B}\right]\pm \frac{160}{7\pi^2}\frac{d}{r_B} \right).
\end{split}
\end{equation*}Notice that up to the first two leading terms, we have the following relations:
\begin{equation*}\begin{split}
E_{\text{Cas}}^{\text{CC},T=0} \sim  E^{\text{DD}, T=0}_{\text{Cas}}+E^{\text{NN}, T=0}_{\text{Cas}},\\
E_{\text{Cas}}^{\text{CP},T=0} \sim E^{\text{ND}, T=0}_{\text{Cas}}+E^{\text{DN}, T=0}_{\text{Cas}}.
\end{split}\end{equation*}The first relation says that the first two leading terms of the zero temperature Casimir energy between two perfectly conducting spheres are the sum of the first two leading terms  of the zero temperature Casimir energy between two Dirchlet spheres and the first two leading terms of the zero temperature Casimir energy between two Neumann spheres. This has been claimed to be true for any two perfectly conducting bodies in \cite{20}.

By taking derivative with respect to $d$, we find that for the zero temperature Casimir force, the first two leading terms are
\begin{equation*}
\begin{split}
F^{\text{DD}, T=0}_{\text{Cas}}\sim & -\frac{\pi^3 r_Ar_B}{720 d^3(r_B\mp r_A)}\left(1\pm \frac{1}{2}\frac{d}{r_B\mp r_A}+\frac{1}{6}\left[\frac{d}{r_A}\mp\frac{d}{r_B}\right]\right),\\
F^{\text{RR}, T=0}\sim &-\frac{\pi^3 r_Ar_B}{720 d^3(r_B\mp r_A)}\left(1\pm \frac{1}{2}\frac{d}{r_B\mp r_A}+\frac{1}{6}\left[\frac{d}{r_A}\mp \frac{d}{r_B}\right]+\frac{10}{ \pi^2}\frac{d}{r_A}(3\alpha_A-2)\mp \frac{10}{\pi^2}\frac{d}{r_B}(3\alpha_B-2)\right),\\
F^{\text{RD}, T=0}\sim & \frac{7\pi^3 r_Ar_B}{5760 d^3(r_B\mp r_A)}\left(1\pm \frac{1}{2}\frac{d}{r_B\mp r_A}+\frac{1}{6}\left[\frac{d}{r_A}\mp\frac{d}{r_B}\right]+\frac{40}{7\pi^2}\frac{d}{r_A}(3\alpha_A-2)\right),\\
F^{\text{DR}, T=0}\sim &\frac{7\pi^3 r_A r_B}{5760 d^3(r_B\mp r_A)}\left(1\pm\frac{1}{2}\frac{d}{r_B\mp r_A}+\frac{1}{6}\left[\frac{d}{r_A}\mp\frac{d}{r_B}\right]\mp\frac{40}{7\pi^2}\frac{d}{r_B}(3\alpha_B-2)\right),\\
F_{\text{Cas}}^{\text{CC},T=0} \sim & -\frac{\pi^3 r_Ar_B}{360 d^3(r_B\mp r_A)}\left(1\pm \frac{1}{2}\frac{d}{r_B\mp r_A}+\left[\frac{1}{6}-\frac{10}{\pi^2}\right]\left[\frac{d}{r_A}\mp \frac{d}{r_B}\right]\right),\\
F_{\text{Cas}}^{\text{CP},T=0} \sim &\frac{7\pi^3 r_Ar_B}{2880 d^3(r_B\mp r_A)}\left(1\pm \frac{1}{2}\frac{d}{r_B\mp r_A}+\left[\frac{1}{6}-\frac{40}{7\pi^2}\right]\left[\frac{d}{r_A}\mp \frac{d}{r_B}\right]\right).
\end{split}
\end{equation*}In the case of CC boundary conditions, the next-to-leading order term has been computed numerically in \cite{21}. It reads as
\begin{equation*}
F_{\text{Cas}}^{\text{CC},T=0} \sim -\frac{\pi^3 r_Ar_B}{360 d^3(r_B\mp r_A)}\left(1\pm \frac{k_1}{2}\frac{d}{r_B\mp r_A}-\frac{k_2}{2}\frac{d}{r_A}\pm  \frac{k_3}{2} \frac{d}{r_B} \right),
\end{equation*}where
\begin{equation*}
\begin{split}
k_1=1.08(\pm 0.08),\quad k_2=1.38(\pm0.06),\quad k_3=1.05(\pm 0.14).
\end{split}
\end{equation*}
Our exact computation gives
\begin{equation*}
F_{\text{Cas}}^{\text{CC},T=0} \sim   -\frac{\pi^3 r_Ar_B}{360 d^2(r_B\mp r_A)}\left(1\pm \frac{1}{2}\frac{d}{r_B\mp r_A}-\frac{1.69}{2}\frac{d}{r_A}\pm \frac{1.69}{2} \frac{d}{r_B} \right).
\end{equation*}One can see that $k_1$ agrees quite well with the exact value $1$, but the errors in the fits for $k_2$ and $k_3$ are about 15\% and 30\% respectively.

Finally, let us consider the case where sphere $B$ is much larger than sphere $A$, i.e., $r_A\ll r_B$. In the case of two cylinders, such scenario has been considered in \cite{23}. Since
\begin{equation*}
\frac{r_B}{r_B\mp r_A}=1\pm\frac{r_A}{r_B}+\ldots,
\end{equation*}we find that the small separation asymptotic expansions of the zero temperature Casimir energies are given by
\begin{equation*}
\begin{split}
E^{\text{DD}, T=0}_{\text{Cas}}\sim & -\frac{\pi^3 r_A }{1440 d^2 }\left(1\pm\frac{r_A}{r_B} +\frac{1}{3} \frac{d}{r_A}\pm\frac{2}{3}\frac{d}{r_B} \right),\\
E^{\text{RR}, T=0}\sim &-\frac{\pi^3 r_A }{1440 d^2 }\left(1\pm \frac{r_A}{r_B } +\left[\frac{1}{3}+\frac{20}{ \pi^2}(3\alpha_A-2)\right]\frac{d}{r_A}\pm\left[\frac{2}{3}- \frac{20}{\pi^2}(3\alpha_B-2)\right]\frac{d}{r_B}\right),\\
E^{\text{RD}, T=0}\sim & \frac{7\pi^3 r_A }{11520 d^2 }\left(1\pm \frac{r_A}{r_B} +\left[\frac{1}{3}+\frac{80}{7\pi^2}(3\alpha_A-2)\right]\frac{d}{r_A}\pm\frac{2}{3}\frac{d}{r_B}\right),\\
E^{\text{DR}, T=0}\sim &\frac{7\pi^3 r_A  }{11520 d^2 }\left(1\pm\frac{r_A}{r_B}+\frac{1}{3} \frac{d}{r_A} \pm\left[\frac{2}{3}-\frac{80}{7\pi^2}(3\alpha_B-2)\right]\frac{d}{r_B}\right),
\\
E_{\text{Cas}}^{\text{CC},T=0} \sim & -\frac{\pi^3 r_A }{720 d^2 }\left(1\pm \frac{r_A}{r_B }+\left[\frac{1}{3}-\frac{20}{\pi^2}\right] \frac{d}{r_A}\pm \left[\frac{2}{3}+\frac{20}{\pi^2}\right] \frac{d}{r_B} \right),\\
E_{\text{Cas}}^{\text{CP},T=0}\sim & \frac{7\pi^3 r_A }{5760 d^2 }\left(1\pm \frac{r_A}{r_B }+\left[\frac{1}{3}-\frac{80}{7\pi^2}\right] \frac{d}{r_A}\pm
\left[\frac{2}{3}+\frac{80}{7\pi^2}\right] \frac{d}{r_B} \right).
\end{split}\end{equation*}
 In the limit $r_B\rightarrow\infty$, we obtain the configuration of a sphere  in front a plane. In this limit,  $d/r_B, r_A/r_B\rightarrow 0$. The results obtained above reproduce the results for a sphere in front of a plane obtained in \cite{22}.

\section{The exact leading term of the free energy}
In this section, we consider the Casimir free energy at finite temperature. We are only going to consider the leading term. As mentioned in Section \ref{s2}, one can obtain the representation for the free energy from the zero temperature Casimir energy by using the Matsubara formalism, which involves changing the integration over the imaginary frequency $\xi$ to summation over the Matsubara frequencies $\xi_p=2\pi p T$.

First consider the scalar case. From \eqref{eq11_30_11}, we find that when sphere $A$ is inside sphere $B$, the leading term of the zero temperature Casimir energy is
\begin{equation}\label{eq12_02_1}\begin{split}
E^{\text{XY}, T=0}_{\text{Cas}}\sim &
 -\frac{b}{4\pi r_A}\sum_{s=0}^{\infty}\frac{(-1)^{(x+y)(s+1)}}{ (s+1)^{2}}\int_0^{1} \frac{\tau^{-1}d\tau}{ \sqrt{1-\tau^2}}\int_{0}^{\infty}dl\,l \exp \left( -\frac{2\vep l(s+1)}{a\tau}\right).
\end{split}\end{equation}Recall that
$$\xi=\frac{\omega}{r_B-r_A}=\frac{l\sqrt{1-\tau^2}}{r_A\tau}.$$Therefore,
\begin{equation*}
\tau=\frac{l}{\sqrt{l^2+[r_A\xi]^2}}.
\end{equation*}Changing $\tau$ back to $\xi$ and replacing $s+1$ with $k$, we have
\begin{equation}\label{eq12_02_2}\begin{split}
E^{\text{XY}, T=0}_{\text{Cas}}\sim &
 -\frac{r_B}{4\pi (r_B-r_A)}\int_{0}^{\infty}d\xi\sum_{k=1}^{\infty}\frac{(-1)^{k(x+y) }}{ k^{2}}\int_0^{\infty}dl \frac{l}{\sqrt{l^2+[r_A\xi]^2}} \exp \left( -\frac{2kd\sqrt{l^2+[r_A\xi]^2}}{r_A}\right).
\end{split}\end{equation}
Therefore, for the free energy, the leading term is
\begin{equation}\label{eq12_02_3}\begin{split}
E^{\text{XY}}_{\text{Cas}}\sim &
 -\frac{r_B T}{2 (r_B-r_A)}\sum_{k=1}^{\infty}\frac{(-1)^{k(x+y) }}{ k^{2}} \sum_{p=0}^{\infty}\!' \int_0^{\infty} dl\frac{l}{\sqrt{l^2+[r_A\xi_p]^2}} \exp \left( -\frac{2kd\sqrt{l^2+[r_A\xi_p]^2}}{r_A}\right).
\end{split}\end{equation}
Using the formula
\begin{equation*}
\frac{1}{\alpha}e^{-2\alpha \beta}=\frac{1}{\sqrt{\pi}}\int_0^{\infty} t^{-\frac{1}{2}-1}\exp\left(-\frac{\alpha^2}{t}-t\beta^2\right)dt,
\end{equation*}
and integrating over $l$, we find that
\begin{equation}\label{eq12_02_6}\begin{split}
E_{\text{Cas}}^{ \text{XY}}
\sim&-\frac{r_BT}{2\sqrt{\pi}r_A(r_B-r_A)}\sum_{p=0}^{\infty}\!'\sum_{k=1}^{\infty} \frac{(- 1)^{k(x+y)}}{k^{2}} \int_{0}^{\infty}dl\,l \int_0^{\infty}t^{-\frac{1}{2}-1}
\exp\left(-td^2k^2 -\frac{l^2+[2\pi p r_AT]^2}{tr_A^2}\right)dt\\
= &-\frac{r_Ar_BT}{4\sqrt{\pi}(r_B-r_A)}\sum_{p=0}^{\infty}\!'\sum_{k=1}^{\infty} \frac{(- 1)^{k(x+y)}}{k^{2}}   \int_0^{\infty}t^{\frac{1}{2}-1}
\exp\left(-t d^2k^2 -\frac{ [2\pi p  T]^2}{t}\right)dt\\
 =&-\frac{r_Ar_BT}{4\sqrt{\pi}(r_B-r_A)}\sum_{k=1}^{\infty} \frac{(- 1)^{k(x+y)}}{k^{2}} \left(\frac{1}{2}\frac{\Gamma\left(\frac{1}{2}\right) }{ kd}+ \sum_{p=1}^{\infty}\frac{\sqrt{\pi}}{kd}e^{-4\pi kpdT}\right)\\
  =&-\frac{r_Ar_BT}{8d(r_B-r_A)}\sum_{k=1}^{\infty}  (- 1)^{k(x+y)}  \frac{\coth 2\pi kdT}{k^3}.
  \end{split}
\end{equation}Comparing to \eqref{eq11_25_3} and \eqref{eq11_25_3_2}, one   sees immediately that this is exactly the proximity force approximation to the Casimir free energy. In particular, the leading terms of the Casimir free energy in the medium and high temperature regions agree with those predicted by the proximity force approximation. For the low temperature region, the leading term is the zero temperature term as predicted by the proximity force approximation. However, our method does not yield the low temperature leading term of the thermal correction to the Casimir free energy. Different methods have to be employed and we would not discuss it here.

For electromagnetic fields, \eqref{eq12_01_4} shows that up to the leading term,
\begin{equation*}
\begin{split}
E_{\text{Cas}}^{ \text{CC}} \sim 2E_{\text{Cas}}^{ \text{DD}},\\
E_{\text{Cas}}^{ \text{CP}} \sim 2E_{\text{Cas}}^{ \text{DR}}.
\end{split}
\end{equation*}These again agree with the proximity force approximations. For the case where the two spheres are outside each other, we also obtain the same results as those obtained by the proximity force approximations.

If we use the method of finding the next-to-leading order term of the zero temperature Casimir energy to find the next-to-leading order of the Casimir free energy at the medium and the high temperature regions, we will encounter some difficulties resulted from the divergences  in the summation over $s$. This  signifies that either we should not Taylor-expand the logarithm or some of the approximations we use do not work in the medium and the high temperature regions. This is a complicated issue and we will not consider it here.

\section{Conclusion}
In this article, we have computed the  small distance  asymptotic expansions of the   Casimir  free energy between two spheres from the exact representation of the Casimir free energy. We consider scalar fields with Dirchlet, Neumann or general Robin boundary conditions, as well as electromagnetic fields with perfectly conducting or infinitely permeable boundary conditions. At zero temperature, we compute the leading and the next-to-leading order terms. The results are summarized in \eqref{eq12_15_1}  and \eqref{eq12_15_2}. From these, we also obtain the asymptotic expansions when one of the spheres becomes large. In the limiting situation where the radius of the larger sphere goes to infinity, we recover the asymptotic expansions for the sphere-plane geometry. At finite temperature, we obtain immediately the leading behavior of the Casimir free energy using the Matsubara formalism. The analytic formula we obtain agrees completely with that obtained using the proximity force approximation. For future work, it will be interesting to consider dielectric spheres.

\begin{acknowledgments}\noindent
We would like to thank M. Bordag for interesting and stimulating discussions. This work is supported by the Ministry of Higher Education of Malaysia  under the FRGS grant FRGS/2/2010/SG/UNIM/02/2.
\end{acknowledgments}


\begin{thebibliography}{10}
\bibitem{1} M. Bordag,   Phys. Rev. D \textbf{73}, 125018 (2006).
\bibitem{2} A. Bulgac, P. Magierski and A. Wirzba,   Phys. Rev. D \textbf{73}, 025007 (2006).

\bibitem{3} T. Emig, R. L. Jaffe, M. Kadar and A. Scardicchio,  Phys. Rev. Lett. \textbf{96}, 080403 (2006).
\bibitem{4} T. Emig, N. Graham, R. L. Jaffe and M. Kardar,  Phys. Rev. Lett. \textbf{99}, 170403 (2007).

\bibitem{5} T. Emig, N. Graham, R. L. Jaffe and M. Kardar,   Phys. Rev. D \textbf{77}, 025005 (2008).
\bibitem{6} T. Emig and R. L. Jaffe,  J. Phys. A: Math. Theor. \textbf{41}, 164001 (2008).
\bibitem{7} O. Kenneth and I. Klich,  Phys. Rev. B \textbf{78}, 014103 (2008).
\bibitem{8} K. A. Milton and J. Wagner,  J. Phys. A: Math. Theor. \textbf{41}, 155402 (2008).

\bibitem{9} S. J. Rahi, T. Emig, N. Graham, R. L. Jaffe, and M. Kardar,   Phys. Rev. D \textbf{80}, 085021 (2009).
\bibitem{16} M. Bordag and V. Nikolaev,  J. Phys. A: Math. Theor. \textbf{41}, 164002 (2008).
\bibitem{17} M. Bordag and V. Nikolaev,   Phys. Rev. D \textbf{81}, 065011 (2010).
\bibitem{22} L. P. Teo, M. Bordag and V. Nikolaev, arXiv: 1110.5100, to appear in Phys. Rev. D.
\bibitem{18} L. P. Teo, Phys. Rev. D \textbf{84}, 065027 (2011).
\bibitem{15} M. Bordag and I. Pirozhenko,  Phys. Rev. D \textbf{81}, 085023 (2010).
\bibitem{26} L. P. Teo, Phys. Rev. D \textbf{84}, 025022 (2011).
\bibitem{24} C. D. Fosco, F. C. Lombardo and F. D. Mazzitelli, Phys. Rev. D \textbf{84}, 105031 (2011).
\bibitem{20} G. Bimonte, T. Emig, R. L. Jaffe and M. Kardar, arXiv: 1110.1082.
\bibitem{25} G. Bimonte, T. Emig and M. Kardar, arXiv: 1112.1366.
\bibitem{21} S. Zaheer, S. J. Rahi, T. Emig and R. L. Jaffe, Phys. Rev. A \textbf{81}, 030502(R) (2010).

\bibitem{12} S. Zaheer, S. J. Rahi, T. Emig and R. L. Jaffe, Phys. Rev. A \textbf{82}, 052507 (2010).
\bibitem{27} P. Rogriguez-Lopez, Phys. Rev. B \textbf{84}, 075431 (2011).

\bibitem{10} D. A. R. Dalvit, F. C. Lombardo, F. D. Mazzitelli and R. Onofrio,  Phys. Rev. A \textbf{74}, 020101(R) (2006).
\bibitem{11} F. D. Mazzitelli, D. A. R. Dalvit and F. C. Lombardo,   New. J. Phys. \textbf{8}, 240 (2006).


\bibitem{13} A. Romeo and A. A. Saharian, J. Phys. A: Math. Gen. \textbf{35}, 1297 (2002).
\bibitem{14} L. P. Teo, JHEP \textbf{0911}, 095 (2009).

\bibitem{19} M. W. Reinsch and J. G. Morehead, J. Math. Phys. \textbf{40}, 4782 (1999).


\bibitem{23} M. Bordag and V. Nikolaev, J. Phys. A: Math. Theor. \textbf{42}, 415203 (2009).


\end{thebibliography}
\end{document}